\documentclass[structabstract]{aa}  

\usepackage{graphicx}
\usepackage{txfonts}

\usepackage[usenames]{color}

\newcommand\pycasso{{\sc p}y{\sc casso}}          	
\newcommand\starlight{{\sc starlight}}          	
\newcommand\adev{\overline\Delta}        		

\begin{document}

\title{Resolving galaxies in time and space: I:}
\subtitle{Applying STARLIGHT to CALIFA data cubes}

\authorrunning{Cid Fernandes et al.}
\titlerunning{Applying STARLIGHT to CALIFA data cubes}

\author{R. Cid Fernandes\inst{1,2},
E. P\'erez\inst{1},
R. Garc\'{\i}a Benito\inst{1},           
R. M. Gonz\'alez Delgado\inst{1},
A. L.\ de Amorim\inst{2}, 
S. F. S\'anchez\inst{1,3},
B. Husemann\inst{4},
J. Falc\'on Barroso\inst{5,6},
P.  S\'anchez-Bl\'azquez\inst{7},
C. J. Walcher\inst{4},
\and
D. Mast\inst{1,3}}

\institute{
Instituto de Astrof\'{\i}sica de Andaluc\'{\i}a (CSIC), P.O. Box 3004, 18080 Granada, Spain\\
\email{cid@astro.ufsc.br}
\and
Departamento de F\'{\i}sica, Universidade Federal de Santa Catarina, P.O. Box 476, 88040-900, Florian\'opolis, SC, Brazil
\and
Centro Astron\'omico Hispano Alem\'an, Calar Alto, (CSIC-MPG), C/Jes\'us Durb\'an Rem\'on 2-2, E-04004 Almer\'{\i}a, Spain
\and
Leibniz-Institut f\"{u}r Astrophysik Potsdam, innoFSPEC Potsdam, An der Sternwarte 16, 14482 Potsdam, Germany
\and
Instituto de Astrof\'{\i}sica de Canarias, V\'{\i}a Lactea s/n, E-38200 La Laguna, Tenerife, Spain
\and
Departamento de Astrof\'{\i}sica, Universidad de La Laguna, E-38205, Tenerife, Spain
\and
Departamento de F\'{\i}sica Te\'orica, Universidad Aut\'onoma de Madrid, 28049 Madrid, Spain
}

\date{Accepted April 16, 2013}

 
\abstract
{}
{Fossil record methods based on spectral synthesis techniques have matured during
the past decade, and their application to integrated galaxy spectra fostered substantial advances on the understanding of galaxies and their evolution. Yet, because of the lack of spatial resolution, these studies are limited to a global view, providing no information about the internal physics of galaxies.}
{Motivated by the CALIFA survey, which is gathering Integral Field Spectroscopy (IFS) over the full optical extent of 600 galaxies, we have developed an end-to-end pipeline which: (i) partitions the observed data cube into Voronoi zones in order to, when necessary and taking due account of correlated errors, increase the signal-to-noise ratio, (ii) extracts rest-framed spectra, including propagated errors and bad-pixel flags, (iii) feeds the spectra into the \starlight\ spectral synthesis code, (iv) packs the results for all galaxy zones into a single FITS or HDF5 file, (v) performs a series
of post-processing operations, including zone-to-pixel image reconstruction and unpacking
the spectral and stellar population properties derived by \starlight\ into
multi-dimensional time, metallicity, and spatial coordinates.
This paper provides an illustrated description of this whole pipeline and its
many products. Using data for the nearby spiral NGC 2916 as a show case, we go
through each of the steps involved, and present a series of ways of visualizing
and analyzing this manifold. These include 2D maps of properties such as the
velocity  field, stellar extinction, mean ages and metallicities, mass surface
densities, star formation rates on different time scales and normalized in
different ways, plus 1D averages in the temporal and spatial dimensions, leading
to evolutionary curves and radial profiles of physical properties. Projections
of the stellar light and mass growth onto radius-age diagrams are introduced as
a means of visualizing galaxy evolution in time and space simultaneously,
something which can also be achieved in 3D with snapshot cuts through the
$(x,y,t)$ cubes.}
{The results provide a vivid illustration of the richness of the
combination of IFS data with spectral synthesis, of the insights on galaxy
physics provided by the variety of diagnostics and semi-empirical constraints
obtained, as well as a glimpse of what is to come from CALIFA and future IFS
surveys.}
{}

\maketitle

\section{Introduction}

Last decade technology has lead to  massive surveys either in  imaging or
spectroscopy modes that have been very successful in providing   information of
the spectral energy distribution or central/integrated spectra for a large
number of galaxies (e.g. 2dF, Folkes et al.\ 1999; SDSS, York et al.\ 2000;
COMBO-17, Bell et al.\ 2004; ALHAMBRA, Moles et al.\ 2008; COSMOS, Ilbert, et al
2009). These data have allowed significant insight on the integrated properties
of galaxies, as well as on the evolution of global quantities such as the mass
assembly (P\'erez-Gonzalez et al.\ 2008) and the star formation history (SFH) of the
universe (Lilly et al.\ 1996; Madau et al.\ 1996). However, these surveys are
limited in either spectral or spatial resolution. Imaging provides spatial
information, but at the expense of coarse spectral coverage (typically broad
band filters),  limiting the amount of information on the stellar populations,
and providing little or no information on the gas (emission lines) nor the
kinematics of the galaxy. Surveys based on integrated spectra, offer a richer
list of diagnostics of the gaseous and stellar components, but lack spatial
resolution and suffer from aperture effects (one spectrum per object, centered
at the nucleus and not covering the full galaxy). Moreover, integrated galaxy
spectra do not allow  to, for example, isolate morphological components (bulge,
disc, bars),  map the effects of mergers,  trace secular processes such as
stellar migration, and other features of galaxy formation and evolution which
can only be observationally tackled with a combination of  imaging and
spectroscopic capabilities.

The near future will see a proliferation of Integral Field Spectroscopy (IFS)
surveys, which will allow a detailed look at physics {\em within} galaxies, as
opposed to the global view offered by integrated light surveys. CALIFA, the
Calar Alto Legacy Integral Field Area survey is a pioneer in this area
(S\'anchez et al. 2012), and others will follow soon, like SAMI (Croom et al
2012), VENGA (Blanc et al.\ 2009), and MaNGA (Bundy et al., in prep). CALIFA is obtaining IFS for 600 nearby galaxies
($0.005 < z< 0.03$) of all types, mapping stellar populations, ionized gas and
their kinematics across the full optical extent of the sources. The number of
spectra to be generated is of the same order of the whole SDSS ($\sim 10^6$).

Clearly, however, these will not be ``just another million galaxy spectra".
Processed through the machinery of fossil record methods, which recover the
history of a stellar system out of the information encoded in its spectrum
(Walcher et al.\ 2011 and references therein), CALIFA  will provide valuable and
hitherto unavailable information on the {\em spatially resolved} SFH of
galaxies. An example of what can be done was presented by
P\'erez et al.\ (2012), where we have used data on  the  first 105 galaxies
observed by CALIFA in conjunction with the spectral synthesis code \starlight\
(Cid Fernandes et al.\ 2005) to study the spatially resolved mass assembly history of galaxies. We find that, for massive galaxies, the inner parts grow faster than the outer ones, a clear signature of inside-out growth, and that the relative growth rate depends on the stellar mass of the galaxy, with a maximum relative growth efficiency for  intermediate masses ($\sim 7 \times 10^{11} M_\odot$).

The application of spectral synthesis methods to IFS datacubes opens new ways of
studying galaxies and their evolution. In practice, however, this apparently
simple extension of work done for spatially integrated spectra involves a series
technical details and methodological issues, as well as challenges in the
visualization and representation of the results.

The  purpose of this paper is to present a detailed account of all steps in the processing of CALIFA $(x,y,\lambda)$ datacubes to the multidimensional products derived from the \starlight\ analysis. We thus focus basically on methodology, leaving the exploration of results for subsequent studies. Paper II (Cid Fernandes et al.\ 2013) presents a thorough study of the uncertainties involved in these products, including the effects of random noise, spectral shape errors, and evolutionary synthesis models used in the analysis. Despite the emphasis on CALIFA and \starlight, the issues discussed here and in Paper II are of direct interest to spectral synthesis analysis of IFS data in general.

The paper is organized as follows. Section \ref{sec:CALIFA} presents the data.
Section \ref{sec:RGB} provides a detailed description of the pre-processing
steps which prepare the reduced data for a full spectral fitting analysis.
Section \ref{sec:PyCASSO} reviews the \starlight\ code and the ingredients used
in our analysis. It also introduces the \pycasso\ pipeline, used to pack and
analyze the results of the synthesis. Our main results are presented in Section
\ref{sec:Results}, which uses the nearby spiral NGC 2916 (galaxy 277 in the
CALIFA mother sample) as a show case. This long section starts with considerations on how
to evaluate the quality of the spectral fits, and then proceeds to the
presentation of the synthesis products in different ways, like 2D maps of the
stellar light and mass, extinction, mean ages and metallicities, and star
formation rates maps on different time-scales, 1D maps on either the spatial or
temporal dimensions, and attempts to visualize the evolution of galaxies in time
and space simultaneously. The emphasis is always on methodological issues, but
the examples naturally illustrate the richness of the multidimensional products
of this analysis. Finally, our results are summarized in Section
\ref{sec:Conclusions}.

\section{Data}

\label{sec:CALIFA}

The  goals, observational strategy and overall properties of the survey were
described in S\'anchez et al.\ (2012).  In summary, CALIFA's mother sample
comprises 939 galaxies in the 0.005--0.03 redshift range, extracted from the
SDSS imaging survey to span the full color-magnitude diagram down to M$_r<-18$
mag, and with diameters selected to match the field of the IFU instrument
($\sim 1 \arcmin$ in diameter). A thorough characterization of the sample will
be presented in Walcher et al.\ (in prep). CALIFA will observe $\sim 600$ galaxies
from this mother sample, applying purely visibility-based criteria at each
observing night. The final sample will be representative of galaxies in the
local universe, statistically significant and well selected.

The data analyzed in this paper were reduced using the CALIFA Pipeline version 1.3c, described in the Data Release 1 article (Husemann et al.\  2013). Each galaxy is observed with PMAS (Roth et al.\ 2005, 2010) in the PPAK mode (Verheijen
et al.\ 2004; Kelz et al.\ 2006) at the 3.5m
telescope of Calar Alto Observatory and with two grating setups: the V500,
covering from $\sim 3700$ to 7500 \AA\ at FWHM $\sim 6$ \AA\ resolution, and the
V1200 ($\sim 3650$--4600 \AA, FWHM $\sim 2.3$ \AA). The wide wavelength coverage
of the V500 is ideal for our purposes, but the blue end of its spectra, which
contains valuable stellar population tracers such as the 4000 \AA\ break and
high order Balmer lines, is affected by vignetting over a non-negligible part of
the field of view. To circumvent this problem we worked with a combination of
the two gratings, whereby the data for $\lambda < 4600$ \AA\ comes from the
V1200 (which does not suffer from this problem at this wavelength range) and the
rest from the V500. The spectra were homogenized to the resolution of the V500.

For each galaxy the pipeline provides a $(x,y,\lambda)$ cube with a
final sampling of $(1^{\prime\prime},1^{\prime\prime},2 {\rm\AA})$. The cube
comprises absolute calibrated flux densities at each spaxel ($F_\lambda$),
corrected by Galactic extincion using the Schlegel, Finkbeiner \& Davies (1998)
maps and the extinction law of Cardelli, Clayton \& Mathis (1989) law with $R_V = 3.1$. 
Besides fluxes, the reduced cubes contain carefully derived errors ($\epsilon_\lambda$), and a bad-pixel flag ($b_\lambda$) which signals unreliable flux entries, due to, say, bad-columns, cosmic rays and other artefacts (see Husemann et al.\ 2013 for details).

\section{Pre-processing steps: From datacubes to STARLIGHT input} 

\label{sec:RGB}

A series of pre-processing steps are applied to the reduced datacubes, with the
general goal of  extracting good quality spectra for a stellar population
analysis with \starlight\ or any other similar code. These are:

\begin{enumerate} 

\item Definition of a spatial mask

\item Outer $S/N$ mask

\item Refinements of the  $b_\lambda$ (flag) and
$\epsilon_\lambda$ (error) spectra

\item Rest-framing 

\item Spatial binning

\item Resampling in $\lambda$

\end{enumerate}

This section describes each of these operations.

\begin{figure}
\centering \includegraphics[width=0.5\textwidth]{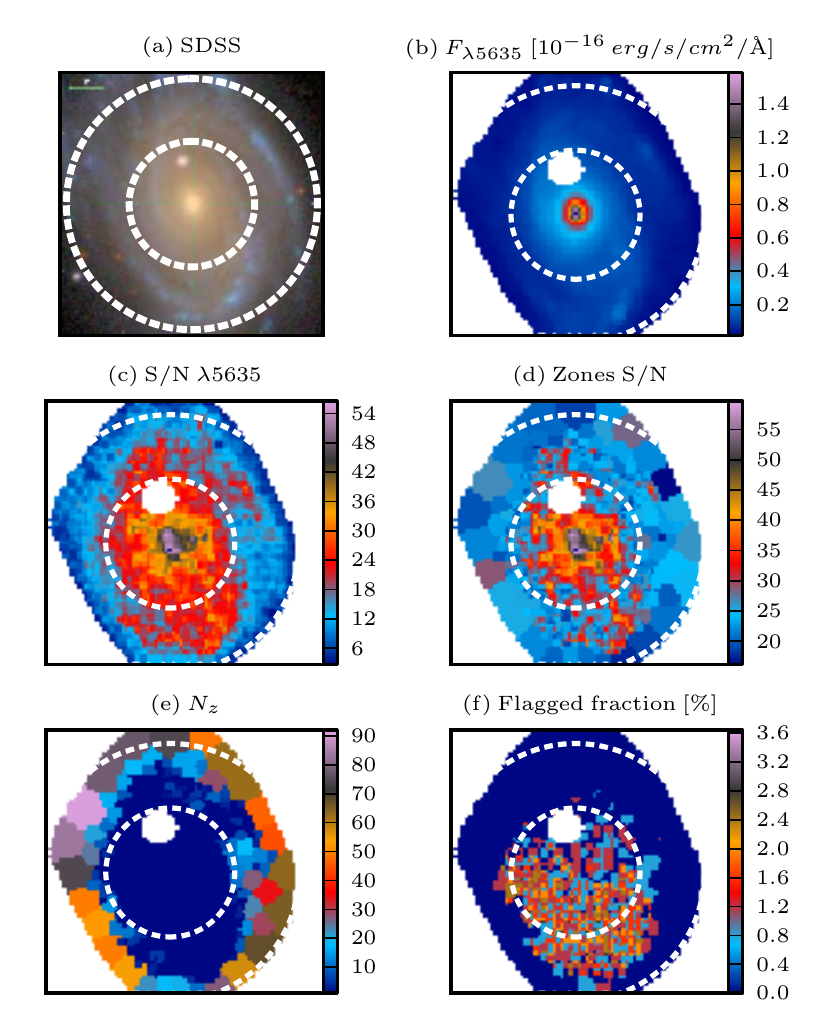}

\caption{(a) SDSS stamp for NGC 2916 (CALIFA 277). As in other images throughout
this paper, the circles mark $R = 1$ and 2 Half Light Radius (HLR). (b) CALIFA
image in the rest-frame $5635 \pm 45$ \AA\ interval, after applying the spatial
mask. (c) Map of the $S/N$ at 5635 \AA. (d) $S/N$ map after the Voronoi binning.
(e) Number of spaxels in each Voronoi zone. Note that no spatial binning ($N_z =
1$) was necessary throughout the inner $\sim 1$ HLR. (f) Percentage of bad
pixels in the 3800--6850 \AA\ range. North is up and east is towards the left in
all  images. }

\label{fig:Q1} 
\end{figure}

\subsection{Spatial masks} 
\label{sec:spatial_masks}

The first step is to mask out foreground stars, artefacts and low $S/N$ regions.
As a starting point, candidate spurious sources are identified applying
SExtractor (Bertin \& Arnouts 1996) to the SDSS $r$-band image. The masked image
is then  matched to the CALIFA field of view using the WCS information available
in the headers to reproject the SDSS masks to the scale, orientation and pixel
size of the CALIFA cubes. To improve upon this first approximation we perform a
visual inspection of each galaxy, correcting its mask interactively whenever
necessary. For instance, we manually  fixed a few unmasked pixels which remained
in the borders of each masked object (probably due to a combination of the
slightly better PSF of the SDSS images and the accuracy of the CALIFA WCS
information used in the matching). Also, in a number of cases we found that the
initial mask included bright H{\sc ii} regions, while in others  very dim
point-like sources can be distinguished in the SDSS image, but have no clear
counterpart in the CALIFA image nor spectra. In all cases, the masks were
corrected by hand.

The resulting masks are free from spurious spaxels. Unlike the other
operations described below, this step could not be fully automated. For the size
of CALIFA, this is still a feasible, albeit laborius, task. A last step in
defining the spatial mask is to impose a spectral quality threshold based on the
$S/N$ of the data. This step  has been fully automated in the
pipeline described next.

\subsection{Flags, errors and segmentation: The {\sc qbick} pipeline}

The next steps (items ii to vi) were all packed in a fully
automated package, named {\sc qbick}, built specifically for CALIFA  but sufficiently general to handle any FITS format to pre-process datacubes for a 3D spectral fitting analysis.

The $b_\lambda$ flags produced by the reduction pipeline ($b_\lambda \equiv 0$
for good pixels and $> 0$ for flagged ones) do a good job in identifying
problematic pixels. For uniformity, we enforce  $b_\lambda > 0$ around the
strongest sky lines seen in the CALIFA data: HgI 4358 \AA, HgI 5461 \AA, OI 5577
\AA, NaI D (around 5890 \AA), OI-OH (6300 \AA), OI 6364 \AA, and the B-band
atmospheric absorption (S\'anchez et al.\ 2007). We have further set  $b_\lambda >
0$ when the value of the error is very large (specifically, when
$\epsilon_\lambda > 5 F_\lambda$), but this is merely a cosmetic choice, as
these pixels would not influence the spectral fits anyway. Anomalously small
$\epsilon_\lambda$ values, on the other hand, can have a disproportionately
large weight  on the analysis. In the very few cases where this happens, the
pixels were flagged. These refinements are useful to circumvent potential
problems in the spectral fitting analysis, specially when the whole data set is
processed in ``pipeline mode".

After these fixes all spectra are rest-framed using the redshift derived by the
reduction pipeline for the central $5^{\prime\prime}$ spectrum (S\'anchez et al.
2012), included in the original headers. This step includes a $(1 + z)^3$
correction for the energy, time and bandwidth changes with redshift.

We used the (rest wavelength) window between 5590 and 5680 \AA\ to estimate a
$S/N$ ratio, defining the signal as the mean flux and the noise as the detrended 
standard deviation\footnote{The detrended std.\ dev.\ is the rms variation with respect to a linear fit, useful to remove the effect of the spectral slope in the window.} of $F_\lambda$, both excluding flagged pixels.\footnote{Note
that this is an apparently superfluous definition, since we already have the
noise spectrum. This "empirical ``$S/N$ is nevertheless useful, as explained in
Section \ref{sec:beta}.} Regions outside the outer $S/N = 5$ iso-contour were
added to the spatial mask, and discarded from the analysis. In a few cases this
threshold was lowered due to low surface brightness.

Fig.\ \ref{fig:Q1}b shows the  $5635 \pm 45$ \AA\ image of  CALIFA 277, after the final spatial mask. The hole above the nucleus corresponds to the foreground source seen in the SDSS image (panel a). The hexagonal pattern of the fiber bundle is clearly visible, since  in this example $S/N > 5$ over most of the field of view (panel c).

The 5635 \AA\ image is also used to compute the galaxy's Half Light Radius (HLR), adding up fluxes in spaxels sorted by distance to the nucleus and picking the radius at which half of the total flux is reached. For CALIFA 277, HLR $= 17.9^{\prime\prime}$. This datacube-based HLR may not be the ideal measure of the size of a galaxy, but it is the most convenient metric for the present analysis for the simple reason that it is based on the exact same data. In Fig.\ \ref{fig:Q1} and all images throughout this paper circles are drawn at $R =1$ and 2 HLR to guide the eye.

The following step (item v in our list) consists of applying a segmentation
structure to the data, grouping spaxels into spatial zones. All spaxels
belonging to the same zone are assigned a common  integer ID label, unique for
each zone. {\sc qbick} stores the  zone-to-$xy$ tensor needed to map zones to pixels
and vice-versa. In this paper we use  spatial binning to increase the $S/N$ for
the spectral analysis. Other applications may prefer to group spaxels according
to, say, morphology or emission line oriented criteria. {\sc qbick} admits an
externally provided segmentation map to allow for such choices.

The zoning scheme adopted here is based on the Voronoi tesselation technique, as
implemented by Cappellari \& Copin (2003), but with modifications to account for
correlated errors (Section \ref{sec:beta}). The code is feed with the signal and
noise images in the $5635 \pm 45$ \AA\ range discussed above. The target $S/N$ is
set to 20. In practice, most individual spaxels inside 1 HLR satisfy this
condition. Fig.\ \ref{fig:Q1}d shows the $S/N$ map of CALIFA 277 after the
spatial binning.  For this galaxy, the  3649 non-masked spaxels were grouped
into 1638 zones, the overwhelming majority (1527) of which are not binned at
all. As seen in the map of the number of spaxels per zone ($N_z$, Fig.\
\ref{fig:Q1}e), spatial binning effectively starts at $R \sim 1$ HLR.
Discounting the single spaxel zones, the median $N_z$ is 10, but values up to
92 are reached in the outskirts.

Once the spatial binning map is defined, the flux, error and flag spectra are
computed for each zone. This operation must account for the flagged pixels. The
total flux in zone $z$, containing $N_z$ spaxels whose fluxes are
$F_{\lambda,k}$, is given by

\begin{equation} 
\label{eq:zone_Flux}
F_{\lambda,z} = 
\frac{N_z}{\Phi_{\lambda,z}} \sum\limits_{k=1}^{N_z} F_{\lambda,k} \phi_{\lambda,k} 
\end{equation}

\noindent where $\phi_{\lambda,k} = 0$ for
flagged pixels ($b_{\lambda,k} > 0$) or 1 otherwise ($b_{\lambda,k} = 0$), and
$\Phi_{\lambda,z} \equiv \sum \phi_{\lambda,k}$ is the number of useful spaxels
at wavelength $\lambda$ in the zone. The zone flux is thus $N_z$ times the
average flux over all non-flagged pixels. When none of the  $F_{\lambda,k}$
fluxes are flagged, this equation simply sums all the fluxes in the zone,
whereas when this is not the case, flagged entries are replaced by the average
of non-flagged ones.

This scaling recipe is reasonable as long as the number of flagged pixels is not
comparable to $N_z$. For example, in a zone of $N_z = 10$ spaxels, one would not
like to take the resulting $F_{\lambda,z}$ seriously if 8 of them are flagged.
To deal with this issue we flag away entries whenever less than half of the spaxels
in a zone have credible data at the given $\lambda$:

\begin{equation} 
b_{\lambda,z} = 
  \left\{ 
  \begin{array}{ll}
         0 & \mbox{if $\Phi_{\lambda,z} \ge N_z / 2$};\\
         1 & \mbox{if $\Phi_{\lambda,z} < N_z / 2$}.
         \end{array} 
	\right. 
\end{equation}

\noindent For our example galaxy, the fraction of flagged pixels in the
3800--6850 \AA\ interval used in the spectral synthesis analysis ranges from 5
to 8\%, with a median of 6\% (see Fig.\ \ref{fig:Q1}f).

The computation of the error in $F_{\lambda,z}$ follows a recipe similar to eq.\
\ref{eq:zone_Flux}, but adding in quadrature

\begin{equation} 
\label{eq:error_lambda}
\epsilon_{\lambda,z}^2 = 
\beta_z^2 \frac{N_z}{\Phi_{\lambda,z}}  \sum\limits_{k=1}^{N_z} \epsilon_{\lambda,k}^2 \phi_{\lambda,k} 
\end{equation}

\noindent where $\beta_z$ accounts for correlated errors (see below).

{\sc qbick} also produces the total spatially integrated spectrum, following these
same prescriptions. This is useful to compare the analysis of the whole to the
sum of the analysis of the parts (which we do in Section
\ref{sec:TheWhole_x_SumOfTheParts}).

Finally, all the spectra are resampled to 2 \AA. The resulting zone spectra are
stored in ASCII and/or FITS files, with flux , error and flag columns necessary
for a careful spectral analysis. All the quality control images produced in this process are packed in a single FITS file along with the technical
parameters adopted in the different steps ($S/N$ clipping, segmentation map,
etc.)

\subsection{Spatial binning and correlated errors}

\label{sec:beta}

\begin{figure}
\includegraphics[width=0.5\textwidth]{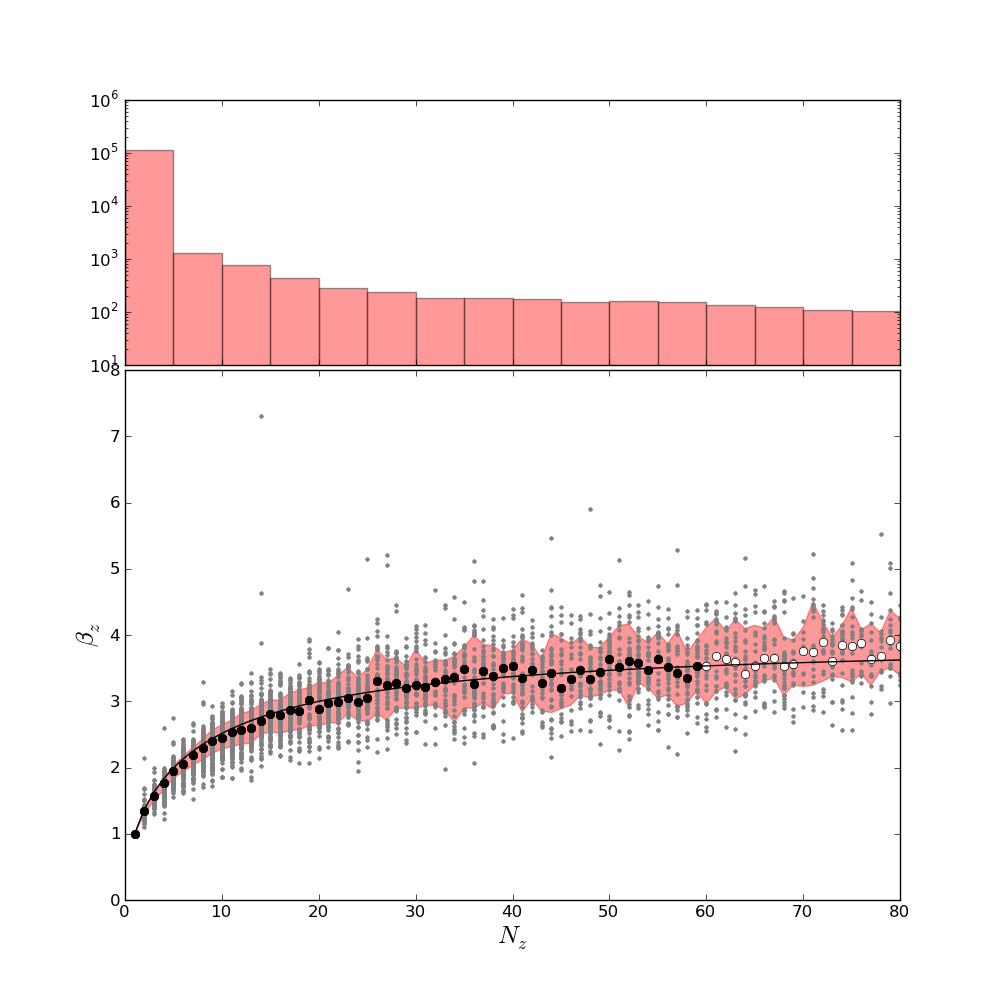}

\caption{Ratio of the measured over the idealized noise for Voronoi zones
comprising $N_z$ spaxels. The idealized noise is that derived under the
assumption of uncorrelated errors, while the measured one is obtained from the
detrended standard deviation of fluxes in a $\pm 45$ \AA\ range around 5635 \AA.
Gray dots represent $\beta_z$ values for 99267 spaxels from 109 galaxies.
Circles show the median $\beta_z$ for each $N_z$, and the solid line shows
equation \ref{eq:beta_fit}. The top panel show the histogram of $N_z$-values
(notice the logarithmic scale).}

\label{fig:beta} 
\end{figure}

As already mentioned, the data cubes have a spatial sampling of
$1^{\prime\prime} \times 1^{\prime\prime}$. Because of the three-fold dithering
pattern used in the observations, each spaxel has a varying contribution from a
number of fibers. Since one fiber can contribute to more than one spaxel, the
noise in a given spaxel is partially correlated with that of its neighbors. This
implies that when spectra from adjacent spaxels are added together the noise in
the resulting spectrum has to be calculated taking spatial covariances into
account. As an illustrative extreme example, consider $N_z$ completely
correlated (ie., $\sim$ identical) spaxels in a zone. Clearly, co-adding their
spectra would not result in the desired increase in the $S/N$, as both signal
and noise scale linearly with $N_z$. Unless told otherwise, however, the zoning
algorithm assumes that the spaxels are independent, and hence that the noise in
the zone is given by $\sqrt{\sum \epsilon_k^2}$, resulting in an improvement of
the $S/N$ of the co-added spectrum by $\sim \sqrt{N_z}$, whereas in fact it has
not changed at all.\footnote{$\epsilon_k$ denotes the noise in spaxel $k$,
defined as the detrended rms in the same $5635 \pm 45$ \AA\ range used to define
the signal for spatial binning purposes.}

An empirical correction was devised to account for the global behaviour of the
spatial correlation of the signal in CALIFA data. The way we proceed is as
follows: First we define initial Voronoi zones with the Cappellari \& Copin (2003) algorithm. Because it assumes uncorrelated input, the code presumes the zone error is $\sqrt{\sum \epsilon_k^2}$. We then measure the ``real'' noise ($e_z$) in
the resulting zone summed spectra from the (detrended) rms in the $5635 \pm 45$
\AA\ range. The ratio of these two numbers, which we call $\beta_z$, is
represented versus the number of spaxels in each  zone in Fig.\ \ref{fig:beta}.
Small grey dots are $\beta_z$ values for the individual zones, circles represent
the median value computed at each $N_z$, while the shaded region represents the
16 to 84 percentiles. There are nearly $10^5$ points in this plot, obtained from
244196 spaxels in 109 galaxies (all those observed with the V500 and V1200
setups up to May 2012). The top panel shows the histogram of $N_z$.

The rise and flattening of the $\beta_z$-$N_z$ relation reflects the expected
behavior. Zones with small $N_z$ group adjacent and highly correlated spaxels,
producing the initial rise of $\beta_z$. As farther apart spaxels are
aggregated, the errors become less correlated, leading to the flattening
observed. The scatter in the relation is likely related to the way the Voronoi
algorithm assemble spaxels. Also, the three-fold dithering pattern is not
completely uniform, so the amount of correlation is not exactly the same across
the face of the fiber bundle. Still, the $\beta_z(N_z)$ relation portrayed in
Fig.\ \ref{fig:beta} is tight enough to derive an excellent statistical
correction.

The solid line represents the fit of the following function:

\begin{equation} 
\label{eq:beta_fit} 
\beta_z(N_z)  
=  \left({{15 N_z}\over{15+N_z-1}}\right)^{1/2} 
\end{equation}

\noindent where the fit was carried out considering only the filled circles. We
can now apply this correction directly in the zoning code, replacing its
idealized computation of the noise by

\begin{equation}
e_{z} = \beta_z(N_z) \sqrt{\sum_{k=1}^{N_z}{\epsilon_k}^2 }
\end{equation}

\noindent which accounts for the effects of spatially correlated errors. This
correction was applied $\lambda$ by $\lambda$ (cf.\ equation \ref{eq:error_lambda}). An independently derived correction very similar to equation \ref{eq:beta_fit} is presented in Husemann et al.\ (2013).

This scheme naturally leads to larger zones than those obtained under the
(erroneous) hypothesis of independent spaxels, but ensures that the target $S/N$
of 20 is actually achieved in all but the outermost zones (when the algorithm
runs out of spaxels to bin). Typically, we obtain 1000 zones per galaxy.

\section{STARLIGHT runs and PyCASSO}

\label{sec:PyCASSO}

\starlight\ is a spectral synthesis method which combines the spectra from a
base in order to match an observed spectrum $O_\lambda$. The search for an
optimal model $M_\lambda$ and the coefficients of the population vector
$\vec{x}$ associated with the $N_\star$ base elements also allows for reddening,
a velocity off-set $v_\star$ and a stellar velocity dispersion $\sigma_\star$.
The code was first introduced in Cid Fernandes et al.\ (2004) and substantially
optimized by Cid Fernandes et al.\ (2005). The code, didactic introductions and a
user manual are available at www.starlight.ufsc.br.
We are actually using a new and yet to be documented version of \starlight, but since most of the new features are not really used in our analysis, the publicly available manual remains an accurate reference for the purposes of this paper.

The virtues and caveats of full spectral fits are discussed in a number of articles (e.g., Panter et al.\ 2003; Ocvirk et al.\ 2006; Cid Fernandes 2007; Tojeiro et al.\ 2007; McArthur, Gonz{\'a}lez \& Courteau 2009; S\'anchez-Bl\'azquez et al.\ 2011). A general consensus in the field is that uncertainties in the results for individual objects average out for large statistical samples (Panter et al.\ 2007). With CALIFA, {\em each galaxy is a statistical sample per se}! Hence, even if the results for single spaxels (or zones) are not iron-clad, the overal trends should be robust
(see Paper II for a detailed evaluation of the uncertainties in our  \starlight-based analysis of CALIFA data.)
It is also worth pointing out that despite the diversity of spectral synthesis methods, substantial changes on the results are more likely to come from revisions in the input data (as happened, for instance, with the recalibration of SDSS spectra between data releases 5 and 6) and from updates in the base models, the single most important ingredient in any spectral synthesis analysis.

The  results reported in this paper rely on a spectral base of Simple Stellar Populations (SSPs) comprising 4 metallicities, $Z = 0.2$, 0.4, 1 and $1,5 Z_\odot$, and 39 ages between $t = 10^6$ and $1.4\times 10^{10}$ yr. This base (dubbed ''{\it GM}'' in Paper II) combines the Granada models of Gonz\'alez Delgado et al.\ (2005) for $t < 63$ Myr with those of Vazdekis et al.\ (2010, as updated by Falc\'on-Barroso et al.\ 2011) for larger ages. They are based on the Salpeter Initial Mass Function and the evolutionary tracks by Girardi et al.\ (2000), except for the youngest ages ($\le 3$ Myr), which are based on Geneva tracks (Schaller et al.\ 1992; Schaerer et al.\ 1993a,b; Charbonnel et al.\ 1993). A comparison of results obtained with other bases is presented in Paper II, to which the reader is also referred for examples of the spectral fits.

The  fits were carried out in the 3800--6850
\AA\ interval. Reddening was modeled with the  Cardelli, Clayton \& Mathis
(1989) curve, wit $R_V = 3.1$. As in Cid Fernandes et al.\ (2005), the main emission lines as well
as the NaI D doublet (because of its sensitivity to ISM absorption) were masked.
All runs were performed in the IAA-GRID, a network of computers which allows us
to process $10^5$ spectra in less than a day.

As documented in the user manual, \starlight\ outputs a large array of quantities. In
broad terms, these can be categorized as

\begin{enumerate}

\item Input data: File names, configuration options, and other information provided either by the user explicitly or derived from the user-provided spectrum. Base-related data are also reported, including ages and metallicities of each component, as well as the corresponding light-to-mass ratios (at the chosen normalization wavelength, $\lambda_N$) and a returned-mass correction factor.

\item Fit results: Figures of merit ($\chi^2$, $\adev$), kinematic parameters ($v_\star$,
$\sigma_\star$), the V-band extinction ($A_V$), total stellar masses and luminosities, and
population vectors, expressed in terms of light ($\vec{x}$) and mass ($\vec{\mu}$)
fractions.

\item Spectral data: The input  ($O_\lambda$),  best model fit $(M_\lambda$),
and  weight spectra ($w_\lambda$, which equals $\epsilon_\lambda^{-1}$ except
for flagged, masked and clipped pixels).

\end{enumerate}

All these data are stored in a plain ASCII file, one for each fitted spectrum.
The population vectors can be post-processed in a number of ways to obtain SFHs,
to compute the mass  growth as a function of time, to condense the results into
first moments like the mean $t$ and $Z$, etc. This scheme works fine for studies
of independent objects, like SDSS galaxies or globular clusters.

IFS data, however, require a more structured level of organization, as it is
cumbersome and inefficient to handle so many files separately. To handle
datacubes we developed \pycasso, the Python Califa Starlight Synthesis
Organizer. \pycasso\ comprises three main parts:

\begin{enumerate}

\item A {\em writer} module, which packs the output of all \starlight\ fits of the individual zones of a same galaxy into a single FITS (or HDF5) file, which also contains all information generated during the pre-processing steps (Section \ref{sec:RGB}), as well as information propagated from the original datacube.

\item A {\em reader} module, which reads this file and structures all the data in an easy to access and manipulate format.

\item A {\em post-processing} module, which performs a series of common operations such as
mapping any property from zones to pixels, resampling and smoothing the population
vectors, computing growth functions, 
averaging in spatial, time or metallicity dimensions,
etc. All these functionalities are conveniently wrapped as python packages, easily
imported into the user's own analysis code.

\end{enumerate}

\pycasso\ will be the main work-horse behind a series of articles by our
collaboration, so we dedicate the remainder of this paper to illustrating  its main products. 

\section{Results}

\label{sec:Results}

The  Sb galaxy NGC 2916 (CALIFA 277) was chosen as a show-case. For the
adopted distance of 56 Mpc, $1^{\prime\prime}$ ($= 1$ CALIFA spaxel) corresponds
to 270 pc. With an  $r$-band absolute magnitude of $-21.17$,  $u-r = 2.18$ and
$g-r=0.54$ colors, and concentration index $C = 2.16$, NGC 2916 can be
characterized as a blue-disk galaxy with an intermediate mass ($\sim 10^{11}
M\odot$). Rudnick, Rix \& Kennicutt (2000) have shown that the
galaxy show some degree of lopsidedness, indicating a weak interaction (possibly
with its irregular satellite $5^\prime$ away, Guti\'errez, Azzaro \& Prada 2002)
or a minor merger. Moorthy \& Holtzman (2006) have obtained long-slit spectra
and measured emission lines and Lick indices along the position angle PA $=
-80^\circ$. They found that the nuclear spectrum is well within the AGN wing of
the BPT diagram [O{\sc iii}]/H$\beta$ vs.\ [N{\sc ii}]/H$\alpha$, in a location
close to what is now accepted as borderline between Seyferts and LINERs (Kewley
et al.\ 2006).

Our Voronoi binning of the CALIFA datacube produces $N = 1638$ zones. The
corresponding $N + 1$ spectra (the $+ 1$ corresponding to the spatially
integrated spectrum) were fitted with \starlight\ and the results feed into
\pycasso. The following subsections describe how we handled the data and the
products of the analysis.

\subsection{Fit quality assessment}

\begin{figure}
\includegraphics[width=0.5\textwidth]{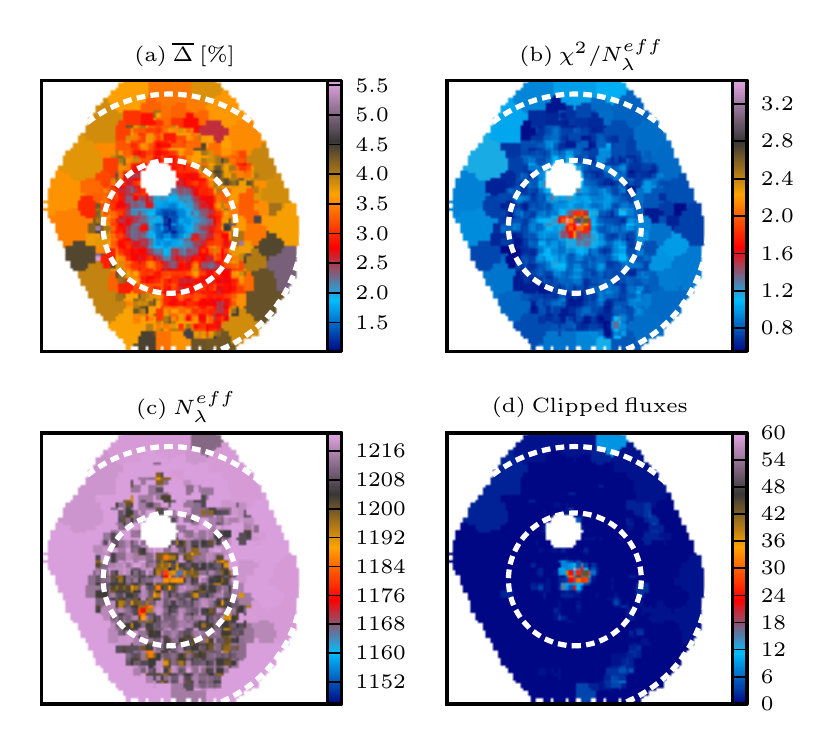}
\caption{Maps of spectral-fit quality indicators. (a) Mean relative deviation
$\adev$, (b) $\chi^2$ per fitted flux, (c) number of fluxes used in the fit, and
(d) number of clipped fluxes.}

\label{fig:adev}
\end{figure}

As with 1D spectra, the quality of the spectral fits should be examined before
proceeding to interpretation of the results. This subsection describes some
standard methods to do this, the caveats involved and strategies to circumvent
them.

Fig.\ \ref{fig:adev}a shows the map of the fit quality indicator

\begin{equation} 
\label{eq:adev}
\adev = \frac{1}{N_\lambda^{\rm eff}} \sum_{\lambda} \frac{|O_\lambda -
M_\lambda|}{M_\lambda},
\end{equation}

\noindent where $O_\lambda$  and $M_\lambda$ are the observed and model spectra,
respectively, and the sum is carried over the $N_\lambda^{\rm eff}$  wavelengths
actually used in the fit, ie, discarding masked, flagged and clipped
pixels.\footnote{Notice that eq.\ \ref{eq:adev} is nearly, but not exactly
identical to what is called ``adev'' in \starlight's output, the difference being
the use of $M_\lambda$ instead of $O_\lambda$ in the denominator. This avoids faulty (but not flagged) $O_\lambda$ entries with very low flux affecting the statistics. This rarely occurs in CALIFA, and when it does the pixels invariably have large errors, and so are irrelevant for the fit, but can
still affect the value  of $\adev$.} For this galaxy $\adev$ spans the range
between 1.0 and 5.6\%, with a median value of 3.0\%. $\adev$ increases towards
the outer regions, where the $S/N$ is smaller (Fig.\ \ref{fig:Q1}c). The $\chi^2
= \sum w_\lambda^2 (O_\lambda - M_\lambda)^2$ map (Fig.\ \ref{fig:adev}b) shows
the opposite behavior, since the errors also increase outwards. Because of its
non-explicit reliance on the (often hard to compute, if not altogether
unavailable) $\epsilon_\lambda$ spectrum, as well as because of its easily
grasped meaning, $\adev$ is a more useful figure of merit to assess fit quality.

Inspection of the highest $\adev$ spectra often reveal non-masked emission lines
or artifacts, like imperfect masking of foreground sources, slightly misplaced
$b_\lambda$ flags, or $O_\lambda$ values which should have been clipped by
\starlight\ but were not because of large $\epsilon_\lambda$. We emphasize that
such bugs are very rare in CALIFA. The median $\adev$ for the $\sim 10^5$
zone spectra analysed so far is just 4\% (corresponding to an equivalent $S/N$
of 25), and in less than 2\% of the cases $\adev$ exceeds 10\%. This high rate
of success is mainly due to the carefully derived errors and flags in the
reduction pipeline and further refined in our pre-processing steps (Section
\ref{sec:RGB}).

One can use such maps to define spatial masks over which the spectral fits
satisfy some quality threshold. Experiments with the whole data set suggest that
a reasonable quality-control limit should be in the $\adev = 8$ to 10\% range.
Not surprisingly, the large residuals tend to be located in the outer regions of
the galaxies. Generally speaking, results beyond 2--3 HLR should be interpreted
with greater care. No $\adev$-based cut is needed for our example galaxy. For
the benefit of \starlight\ users, it is nevertheless worth making some further
remarks on quality control.

\subsubsection{Subtleties and caveats on quality control}

In full spectral fitting methods,  seemingly trivial operations like imposing a
fit-quality threshold often hide subtleties which can easily go unnoticed, particularly when processing tons of data. The paragraphs below (focused on \starlight\ but extensible to other codes) review some of these.

First, one should distinguish cases where $\adev$ is large because of bad
data from those where large residuals arise because of the user's failure to
properly inform, through spectral masks ($m_\lambda$) or flags ($b_\lambda$), spectral regions which should be ignored in the fit. For instance, one can have an
excellent spectrum of an HII region with several strong emission lines besides
those included in a generic emission line mask feed into \starlight, causing an
artificially large $\adev$. Large velocity offsets can have a similar effect, as
emission lines get shifted out of fixed $m_\lambda$ windows. Similarly, sky
residuals and other artefacts missed out by the $b_\lambda$ flags lead to large
$\adev$ even when the overall spectrum is good. In short, not everything which
fails a blind quality control is actually bad.

The clipping options implemented in \starlight\ help spotting pixels which are
too hard to fit and thus  probably represent non-stellar or spurious features.
Our fits use  the ``NSIGMA'' clipping method and a conservative $4 \sigma$
threshold, meaning that we only clip pixels when $|O_\lambda - M_\lambda| > 4
\epsilon_\lambda$. Fig.\ \ref{fig:adev}c maps the number of clipped points in
CALIFA 277. One sees a peak  in the central pixels, where the $S/N$ is so high
that our $4\sigma$ clipping does not forgive even small $O_\lambda - M_\lambda$
deviations (most often associated with problems with the base models rather than
with the data). Elsewhere, very few pixels were clipped. The point to highlight
here is that clipping only worked because we have a reliable $\epsilon_\lambda$.
Other clipping methods can be used when this is not the case,  but in all cases
the user should always check carefully the output, as it may easily happen that
too many points are clipped. 

Custom-made spectral masks also help. For instance, since Mateus et al.\ (2006),
the \starlight\ analysis of SDSS spectra employs individual masks constructed by
searching for emission lines in the $O_\lambda - M_\lambda$ residual spectrum
obtained from a first  fit, and taking into consideration the local
noise level. $O_\lambda$  is then refitted with this tailor-made $m_\lambda$,
circumventing some of the issues above. This refinement has not yet been
implemented for CALIFA.

Errors, flags and masks are, of course, secondary actors in any spectral
synthesis analysis, but these general remarks illustrate that it pays off to
dedicate them some  attention.

\subsection{2D maps: Stellar light, mass and extinction}

\begin{figure}
\includegraphics[width=0.5\textwidth]{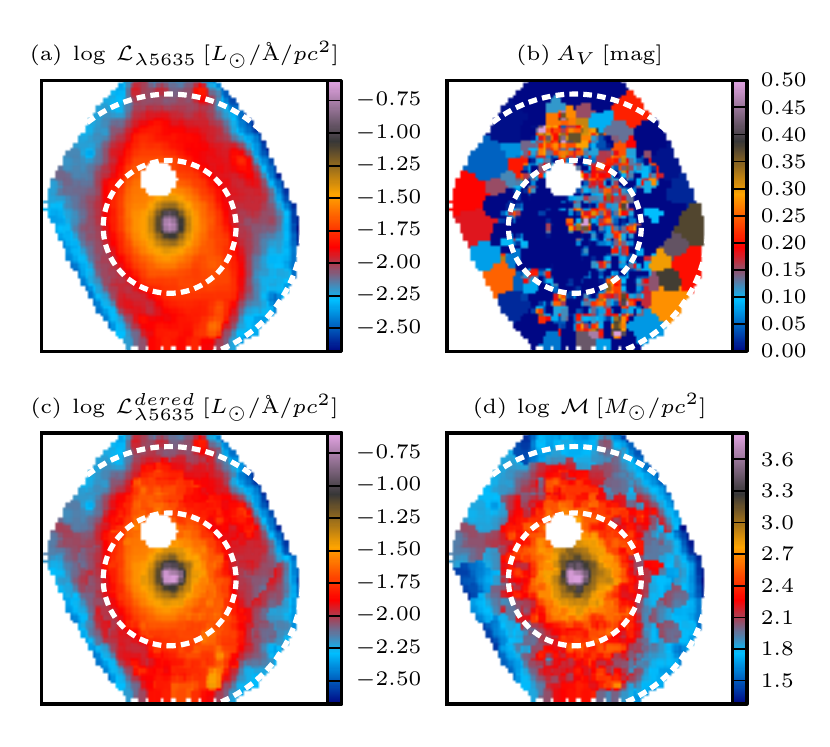}

\caption{(a) Synthesitc surface brightness at 5635 \AA, (b) stellar extinction
map ($A_V$), (c) dereddened ${\cal L}_{\lambda5635}$ image, (d) stellar mass
surface density.}

\label{fig:SBs} 
\end{figure}

One of the main products of spectral synthesis is to convert light to mass.
Fig.\ \ref{fig:SBs} illustrates the results for CALIFA 277. Its top-left panel
shows the surface density of the luminosity  at $\lambda_N = 5635$ \AA, while
the top-right panel show the derived extinction map. The dust corrected  image
and the mass surface density are also shown.

The effects of the Voronoi binning are present in all panels, but are much more
salient in the $A_V$ map. This is because the light and mass images were {\em
``dezonified''} by scaling the value at each $xy$ spaxel by its fractional
contribution to the total flux in a zone ($z$). For instance, for the mass
surface density (${\cal M}$) this operation reads

\begin{equation} 
\label{eq:Mcor_xy}
{\cal M}_{xy} = \frac{M_{xy}}{A_{xy}} = \frac{M_z}{A_z} \times w_{xyz}
\end{equation}

\noindent where $A_{xy}$ ($A_z$) denotes the area in a spaxel (zone), and 

\begin{equation} 
\label{eq:Dezonification}
w_{xyz} = \frac{F_{xy}}{\sum_{xy|z}  F_{xy}} 
\end{equation}

\noindent with $F_{xy}$ as the mean  flux in the $5635 \pm 45$ \AA\ region, the
same used in the Voronoi zoning (Section \ref{sec:RGB}). This operation was
applied to luminosity and mass related quantities, producing somewhat smoother
images than obtained with $w_{xyz} = 1$. Intensive properties (like $A_V$, mean
ages, $\sigma_\star$ and etc.), however, cannot be dezonified.

The stellar extinction map shows low values of $A_V$ (of order 0.1--0.2 mag), with slight enhancements in the nuclear region and the arms. Overall, however, there is relatively little variation across the face of the galaxy (see also Fig.\ \ref{fig:radialProfiles}). The  light and mass images have a similar structure, both showing the bulge at $R \la 0.5$ HLR, and the disc beyond that. The spiral arms are  less prominent in terms of mass than in light because of the higher $L/M$ of stars in the arms.

The total stellar mass obtained from the sum of the zone masses is $M = 6.5 \times 10^{10} M_\odot$. This is the mass locked in stars nowadays. Counting also the mass returned by stars to the interstellar medium, $M^\prime = 9.0 \times 10^{10} M_\odot$ were involved in star formation.These values ignore the mass in the masked region around the foreground star northeast of the nucleus. \pycasso\ can fill in such holes with values estimated from (circular or elliptical) radial profiles. For CALIFA 277, this correction increases the masses quote above by 4\%.

\subsection{2D maps: Kinematics}

\label{sec:kinematics}

\begin{figure}
\includegraphics[width=0.5\textwidth]{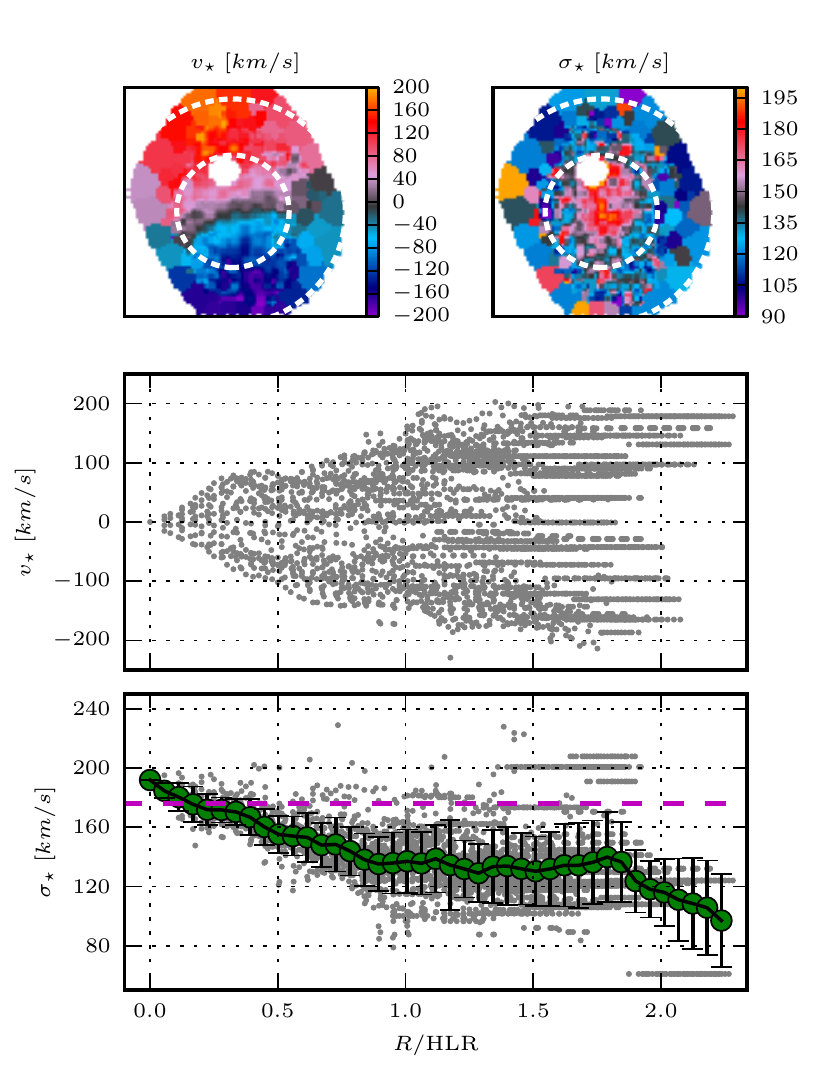}

\caption{Kinematical products of the spectral fitting. The $\sigma_\star$ values are {\em not} corrected for instrumental broadening, which dominates below 140 km/s. The horizontal stripes in the position-velocity and $\sigma_\star(R)$ panels correspond to Voronoi zones.}

\label{fig:kinematics} 
\end{figure}

Fig.\ \ref{fig:kinematics} shows the $v_\star$ and $\sigma_\star$ fields
estimated by \starlight, as well as a position-velocity diagram and a radially
averaged $\sigma_\star$ profile. The $v_\star$ field indicates a projected
rotation velocity of $\sim 200$ km/s. The V500 resolution prevents the
determination of $\sigma_\star$ values below $\sim 150$ km/s, and the  flat  $\sigma_\star(R)$ profile beyond $\sim 0.7$ HLR reflects this resolution
limit (notice that we  have not corrected the values for the instrumental
resolution in this plot). Only in the inner regions the $\sigma_\star$ values
are trustable, reaching 200 km/s in the nucleus (equivalent to 140 km/s after
correcting for instrumental broadening).

The kinematical information derived from our analysis will be superseded by
studies based on the higher spectral resolution V1200 datacubes
(Falc\'on-Barroso et al, in prep). Eventually, one can envisage feeding the parameters derived from these more precise analysis back into the \starlight\ fits (using its fixed-kinematics mode). In fact, this feedback may well turn out
to improve our stellar population analysis, given the potential degeneracy
between $\sigma_\star$ and $Z$ (decreasing the former while increasing the
latter have the same global effect of making  metal lines deeper; Koleva et al
2009; S\'anchez-Bl\'azquez et al.\ 2011).

\subsection{2D maps: Mean ages and metallicities}

\label{sec:mean_ts_and_Zs}

\begin{figure}
\includegraphics[width=0.5\textwidth]{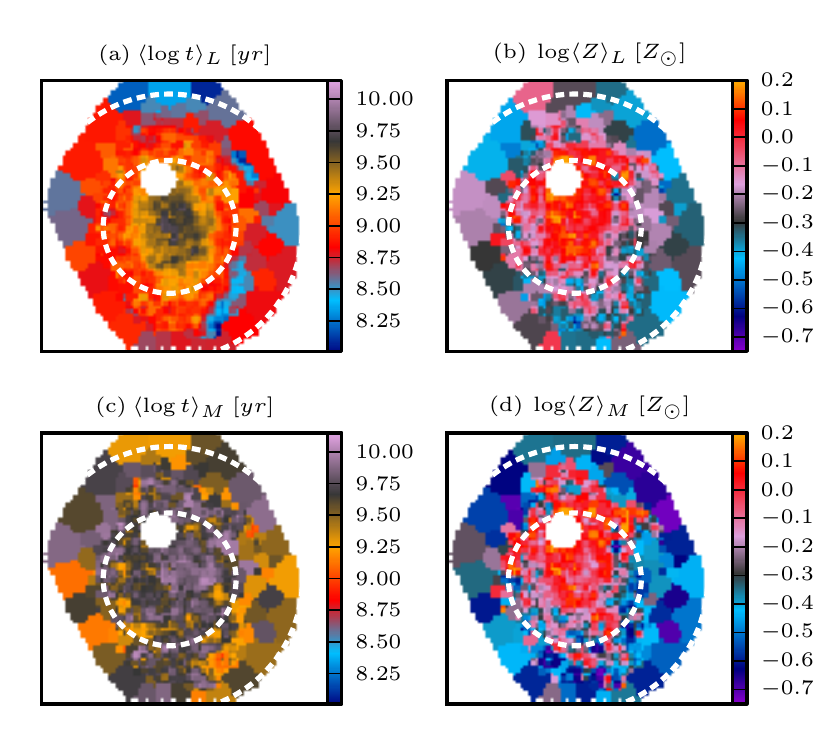}
\caption{Luminosity (top) and mass (bottom) weighted mean age (left) and metallicity (right) maps for CALIFA 277. }
\label{fig:MeanAgeAndZ} 
\end{figure}

The crudest way to quantify the SFH of a system is to compress the age and metallicity
distributions  encoded in the population vectors to their first moments. For this purpose
we will use the following definitions:

\begin{equation} 
\label{eq:at_flux}
\langle \log t\rangle_L = \sum_{t,Z} x_{tZ} \log t 
\end{equation}

\begin{equation}
\label{eq:aZ_flux}
\langle Z \rangle_L = \sum_{t,Z} x_{tZ} Z 
\end{equation}

\noindent where $x_{tZ}$ is the fraction of light at $\lambda_N$ due to the base
population with age $t$ and metallicity $Z$. The mass weighted versions of these
indicators, $\langle \log t\rangle_M$ and $\langle Z \rangle_M$, are obtained replacing $x_{tZ}$ by the mass fraction $\mu_{tZ}$.

Fig.\ \ref{fig:MeanAgeAndZ} shows the light and mass weighted mean (log) age and
metallicity maps. The $\langle \log t\rangle_L$ image shows a steady increase
towards the center. Outside 1 HLR, traces of the spiral arms are noticeable as
regions of lower age (as in the SDSS color image in Fig.\ \ref{fig:Q1}a, the
arms are brighter the western half of the image). Because of the large weight of
old populations, $\langle \log t \rangle_M$ spans a smaller dynamical range than
$\langle \log t\rangle_L$, hence producing lower contrast maps, but the
outwardly decreasing  age is still visible. Negative gradients are also clearly
present in metallicity, with indications of flattening within the bulge region.
As in other 2D maps, the small scale fluctuations towards the edges are at least
in part due to the lower $S/N$.

\begin{figure}
\includegraphics[width=0.5\textwidth]{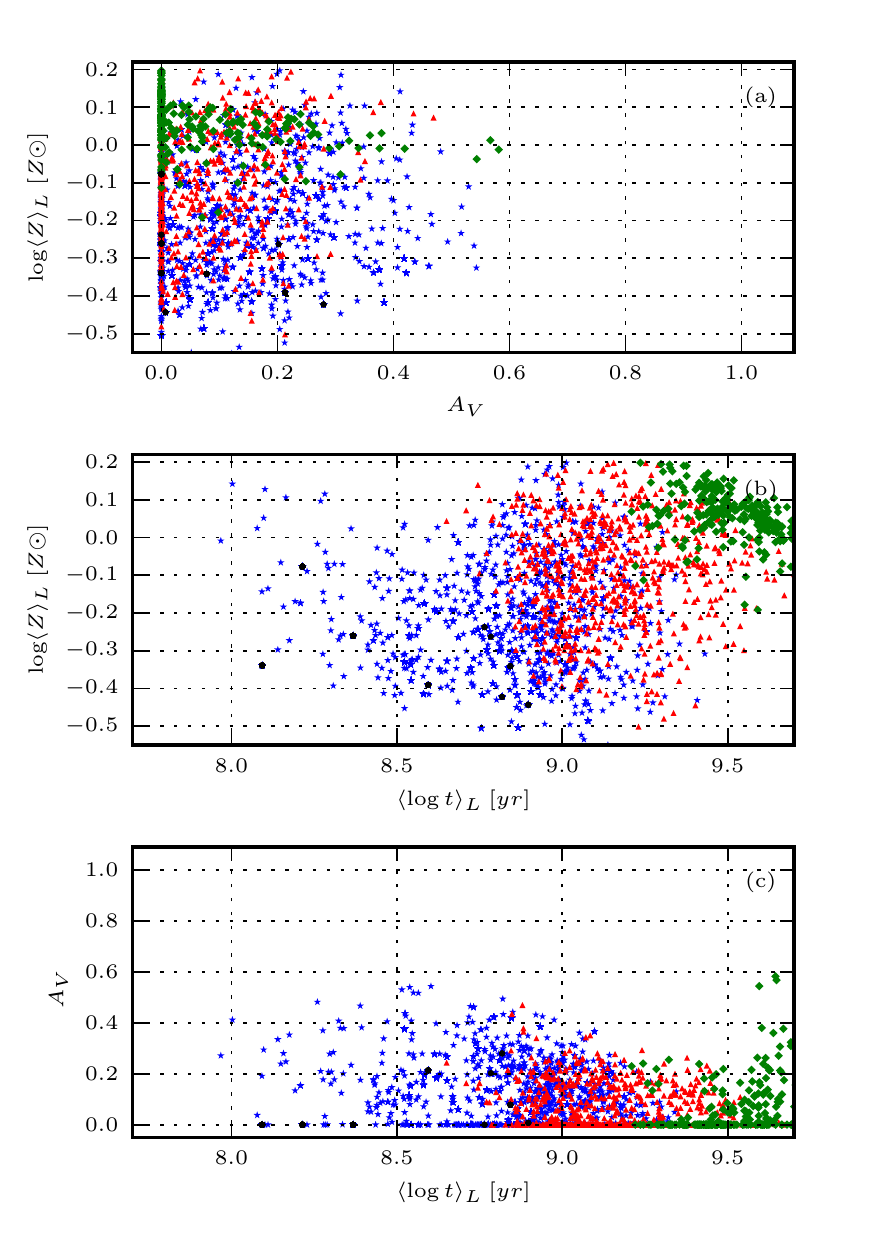}
\caption{Mean age versus mean metallicity versus extinction,
coding zones by their distance to the nucleus. Green diamonds: $R < 0.5$ HLR. Red triangles: 0.5--1 HLR. Blue stars: 1--2 HLR. Black dots: $> 2$ HLR.}
\label{fig:t_x_Z_x_AV} 
\end{figure}

The fact that $t$ and $Z$ grow in the same sense suggests that the
results are not badly affected by the age-metallicity degeneracy. As shown by S\'anchez-Bl\'azquez et al.\ (2011), full spectral synthesis is much less sensitive to this than conventional index-based approaches. Fig.\
\ref{fig:t_x_Z_x_AV} plots extinction versus mean age versus metallicity, the main properties involved in spectral synthesis. Points are color and symbol coded by their distance from the nucleus. One sees that the highest values of $A_V$ are found for old central ($R < 0.5$ HLR) and young outer ($R > 1$ HLR) populations, and that $A_V$ bears no correlation with $Z$. The middle panel shows the positive $t$-$Z$ relation inferred from the 2D maps. Within 0.5 HLR (green diamonds), however, $t$ anticorrelates with $Z$, most likely due to the  age-metallicity degeneracy. The mean $Z$ values in this central region straddle 1 and 1.5 $Z_\odot$, the two highest metallicities in our base.

\subsection{2D maps: $x_Y$, $x_I$, $x_O$}

\label{sec:xYxIxO}

\begin{figure}
\includegraphics[width=0.5\textwidth]{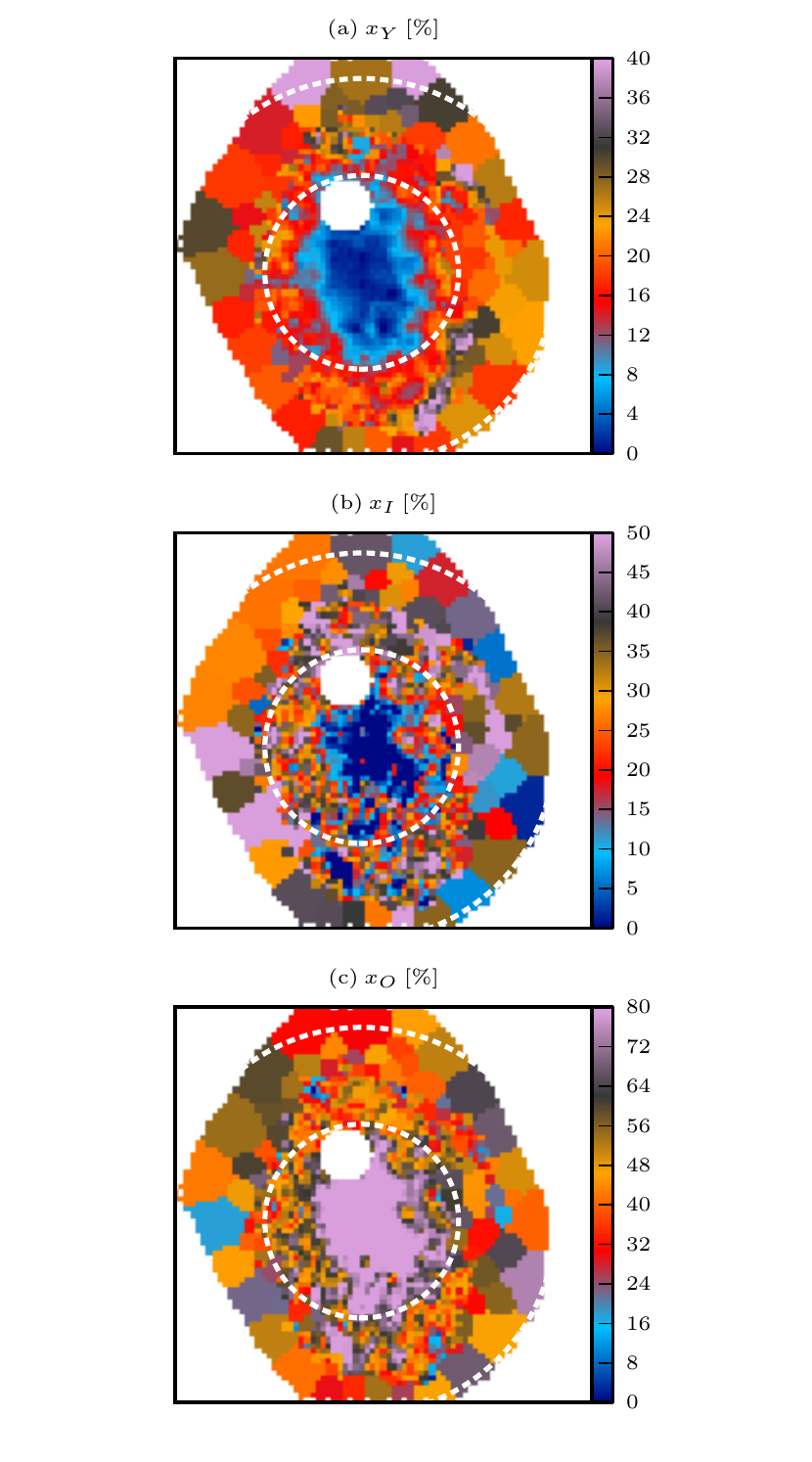}
\caption{Maps of the percentage contribution of Young ($t < 0.14$ Gyr), Intermediate age (0.14--1.4 Gyr), and Old ($> 1.4$ Gyr) stars to the observed light at 5635 \AA.}
\label{fig:xYxIxO} 
\end{figure}

Population synthesis studies in the past found that a useful way to summarize
the SFH  is to condense the age distribution encoded in the population vector
into age ranges. This strategy comes from a time when the analysis was based on
equivalent widths and colors (Bica 1988; Bica et al.\ 1994; Cid Fernandes et
al 2001, 2003), but it  was also  applied to full spectral fits (Gonzalez
Delgado et al.\ 2004).

Fig.\ \ref{fig:xYxIxO} presents images of the light fraction due to Young,
Intermediate and Old populations ($x_Y$, $x_I$ and $x_O$), defined as those with
$t \le 0.14$, $0.14 < t \le 1.4$, and $t > 1.4$ Gyr, respectively. As usual, the
choice of borderlines is somewhat subjective, constrained only by the underlying idea of grouping base elements covering relatively wide age ranges. For
the base used in our \starlight\ runs the Y, I and O bins contain $4 \times 20$,
$4 \times 10$ and $4 \times 9$ SSPs, respectively (where the $4 \times$ comes
from the 4 metallicities).

The plots show that the spiral arms stand out more clearly in the $x_Y$ map,
specially in the western half of the galaxy, as also seen in the SDSS color
composite (Fig.\ \ref{fig:Q1}a). Very little of the light from $R < 1$ HLR comes
from young stars. Intermediate age populations do contribute more, but the
central $\sim 0.5$ HLR is completely dominated by old stars. Overall, however,
despite the added informational content, these maps do not visibly add much to
the $\langle \log t\rangle_L$ image (Fig.\ \ref{fig:MeanAgeAndZ}).

\subsection{2D maps: Star formation rates and IFS-based variations over Scalo's $b$
parameter}

\label{sec:bScalo}

\begin{figure}
\includegraphics[width=0.5\textwidth]{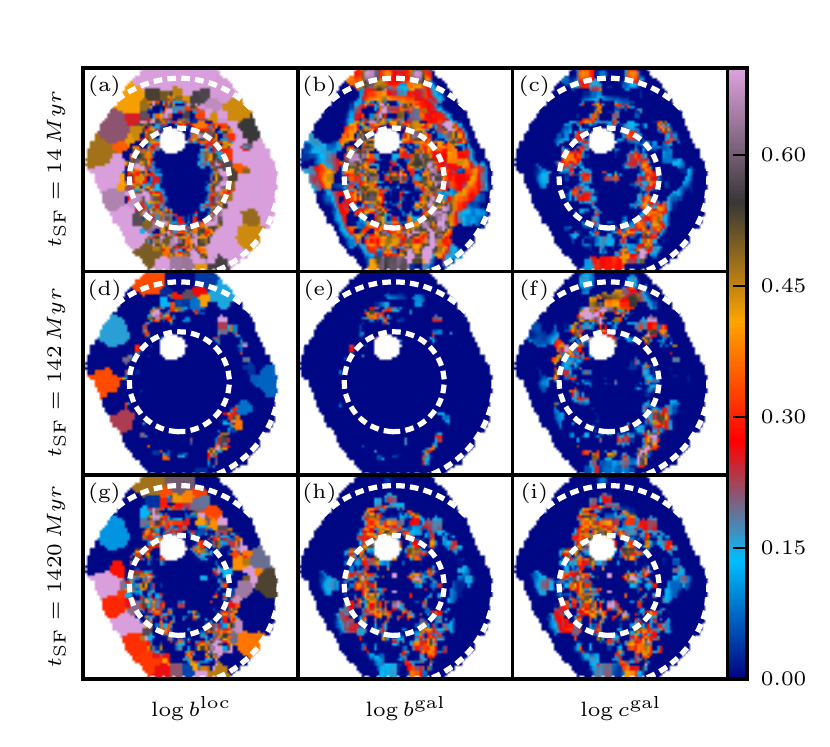}

\caption{Spatially resolved SFR surface density (eq.\ \ref{eq:MoT}) on the last
$t_{SF} = 14$ (top), 142 (middle) and 1420 (bottom) Myr, in units of different
reference values. The left panels compare $\overline{{\cal SFR}}_{xy}(t_{SF})$
to $\overline{{\cal SFR}}_{xy}(t_\infty)$, i.e, the all-times average at $xy$,
thus providing a local version of Scalo's $b$ parameter (eq.\ \ref{eq:b_Loc}).
Middle panels compares the local $\overline{{\cal SFR}}$ to the all-times
average over the whole galaxy (eq.\ \ref{eq:b_Reg}). Panels on the right compare  $\overline{{\cal SFR}}$ to that of the galaxy as a whole over {\em the
same time scale} (eq.\ \ref{eq:c_Reg}). All images are on log scale from
$\log 1$ to $\log 5$, such that only $\overline{{\cal SFR}}_{xy}(t_{SF})$ values above the corresponding reference value are visible.}
\label{fig:bScalo} 
\end{figure}

The base used in our fits comprises instantaneous bursts (ie., SSPs), so,
despite the large number of ages considered, our SFHs are not continuous and
hence not derivable. There are, nonetheless, ways of defining star formation
rates (SFR).

The simplest way to estimate a SFR is to cumulate all the stellar mass formed
since a lookback time of $t_{SF}$ and perform a mass-over-time average:

\begin{equation} 
\label{eq:MoT} 
\overline{{\cal SFR}}_{xy}(t_{SF}) = 
\frac{1}{t_{SF}} \sum_{t < t_{SF}} {\cal M}^\prime_{txy},
\end{equation}

\noindent where ${\cal M}^\prime_{txy}$ is the mass (at $xy$ and per unit area)
of stars formed at lookback time $t$.\footnote{ The stellar mass in this
equation differs from that in eq.\ \ref{eq:Mcor_xy}, Fig.\ \ref{fig:SBs}, and
the definitions of $\langle \log t\rangle_M$ and $\langle Z \rangle_M$. The
latter are corrected by the mass returned to the medium by stars during their
evolution (thus reflecting the mass currently in stars), while in the
computation of SFR's this correction should not be  applied. We distinguish the
two types of stellar masses with the prime superscript (${\cal M}$ versus ${\cal
M}^\prime$).} Eq.\ \ref{eq:MoT} gives the {\em mean} SFR surface density since
$t = t_{SF}$. One can tune $t_{SF}$ to reach different depths in the past, but
as $t_{SF}$ increases this estimator becomes increasingly useless, converging to
the mass density divided by $t_\infty \equiv 14$ Gyr (the largest age in the
base).

It is often more useful to consider SFRs in relation to some fiducial value,
instead of absolute units. The classical example is Scalo's birthrate parameter,
$b$, which measures the SFR in the recent past ($t < t_{SF}$) with respect to
its average over the whole lifetime of the system.\footnote{A related index is
the so called specific SFR, which differs from $b$ just by the $t_\infty$ factor
in equation \ref{eq:b_Loc}.} This is trivially obtained dividing
$\overline{{\cal SFR}}_{xy}(t_{SF})$ by its asymptotic value

\begin{equation} 
\label{eq:b_Loc} 
b^{loc}_{xy}(t_{SF}) = 
\frac{\overline{{\cal SFR}}_{xy}(t_{SF})}{\overline{{\cal SFR}}_{xy}(t_\infty)} = 
\frac{t_\infty}{t_{SF}} \frac{\sum_{t < t_{SF}} {\cal M}^\prime_{txy}}{{\cal M}^\prime_{xy}} 
\end{equation}

\noindent The superscript ``loc'' is to emphasize that the reference lifetime
average SFR is that of same spatial location $xy$. One should recall that while
young stars in a spaxel were probably born there (or near it), old ones may have
migrated from different regions. Thus, despite the identical definitions, the
physical meaning of $b^{loc}_{xy}$ is not really the same as for galaxies as
a whole.

IFS data allow for other definitions of $b$. For instance, one might prefer to compare $\overline{{\cal SFR}}_{xy}(t_{SF})$ to the $\overline{{\cal SFR}}(t_\infty)$ of the galaxy as a whole, the bulge, the disc or any general region. This variation measures the ``present'' ($t < t_{SF}$) ``here'' (at $xy$) to the ``past'' ($t \le t_\infty$) in a spatial region ${\cal R}$:

\begin{equation} 
\label{eq:b_Reg} 
b^{\cal R}_{xy}(t_{SF}) = 
\frac{\overline{{\cal SFR}}_{xy}(t_{SF})}{\overline{{\cal SFR}}_{\cal R}(t_\infty)} 
\end{equation}

A formally similar but conceptually different definition is obtained using for the reference value in the denominator the SFR surface density within the {\em same time-scale} but evaluated in a {\em different region}:

\begin{equation} 
\label{eq:c_Reg} 
c^{\cal R}_{xy}(t_{SF}) = 
\frac{\overline{{\cal SFR}}_{xy}(t_{SF})}{\overline{{\cal SFR}}_{\cal R}(t_{SF})},
\end{equation}

\noindent  i.e., to compare not ``now versus then'', but ``now here versus now
there''. Together with equations \ref{eq:b_Loc} and \ref{eq:b_Reg}, this
definition cover the different combinations of time and space enabled by the
application of fossil methods to IFS data.

Fig.\ \ref{fig:bScalo}  shows maps of $b^{loc}$ (left panels), $b^{\cal R}$
(middle), and $c^{\cal R}$ (right) for three values of $t_{SF}$: 14 (top), 142
(middle) and 1420 Myr (bottom). In all panels the color scale is deliberately
saturated to highlight regions where $\overline{{\cal SFR}}(t_{SF})$ is larger
then chosen reference value---the dynamical range of the images goes from 1 to 5
(0 to 0.7 in log) in the corresponding relative units.

Panel a shows that spaxels which have formed stars over the past 14 Myr at a
larger rate than their respective past average are all located in the outer
regions. Nowhere within $R \la 0.5$ HLR one finds $b^{loc}(14 {\rm Myr}) > 1$.
Considering the past 142 Myr (panel d), the inner "deficit" covers as much as
1.2 HLR. On the longest time scale considered (g) one again sees that the inner
spaxels have been less active than their life-long average.

Maps look considerably different in the middle column of panels, where the local
$\overline{{\cal SFR}}$ is normalized to the past average over the {\em whole
galaxy} (eq.\ \ref{eq:b_Reg}). The spiral arms of CALIFA 277 appear better
delineated in $b^{gal}(14 {\rm Myr})$ (panel b) than in any $b^{loc}$ map.
Interestingly, the $b^{gal}$ map is practically featureless for $t_{SF} = 142$
Myr (e). This suggests that star-formation is not continuous over this time
scale, but instead happens in a bursty mode. Over the past 1.4 Gyr (h), one
again sees traces of the arms, but with a lower amplitude than in panel b,
qualitatively consistent with the cumulative effect of an intermittent sequence
of  short duration bursts. (Due to the logarithmic age resolution of fossil
methods, such short bursts can only be recognized as such at the very young ages
sampled in panel b.)

The right column panels show our novel relative SFR index $c$ (eq.\
\ref{eq:c_Reg}), using the whole galaxy as reference region. Unlike Scalo's $b$,
this index does not compare present to past, but present to present elsewhere.
Its 2D maps can be understood as ``snapshots'' of the SFR in the galaxy taken with
an ``exposure time'' $t_{SF}$. Keeping the ``diaphragm'' open for only the last 14
Myr (panel c) highlights the ongoing star-formation in the galaxy. The spiral
arms are clearly visible, and the structures are more focused than in
the $b^{gal}(14 {\rm Myr})$ map (panel b). Integrating for 142 Myr (f),  the
inner parts of the arms fade, but their outer ($R > 1$ HLR) parts brighten up in
terms of $c^{gal}$. Over the past 142 Myr, these regions have formed more stars
than anywhere else in the galaxy, even though comparing with panel e we find
that only parts of the $c^{gal} > 1$ regions are forming stars at a rate per
unit area larger than that of the galaxy as a whole through its entire life (ie,
$b^{gal} > 1$). For $t_{SF} = 1.4$ Gyr (panel i) the image becomes more blurred.
Because of the long time scale, the $c^{gal}(1.4 {\rm Gyr})$ map is nearly
indistinguishable from that obtained with $b^{gal}$ for the same $t_{SF}$ (panel
h).

One thus sees that, despite some degree of redundancy (the numerator is always
the same), these definitions offer different and complementary views of the star
formation in a galaxy.

\subsection{1D spatial maps: radial profiles}

\label{sec:radialProfiles}

\begin{figure*}
\includegraphics[width=1.0\textwidth]{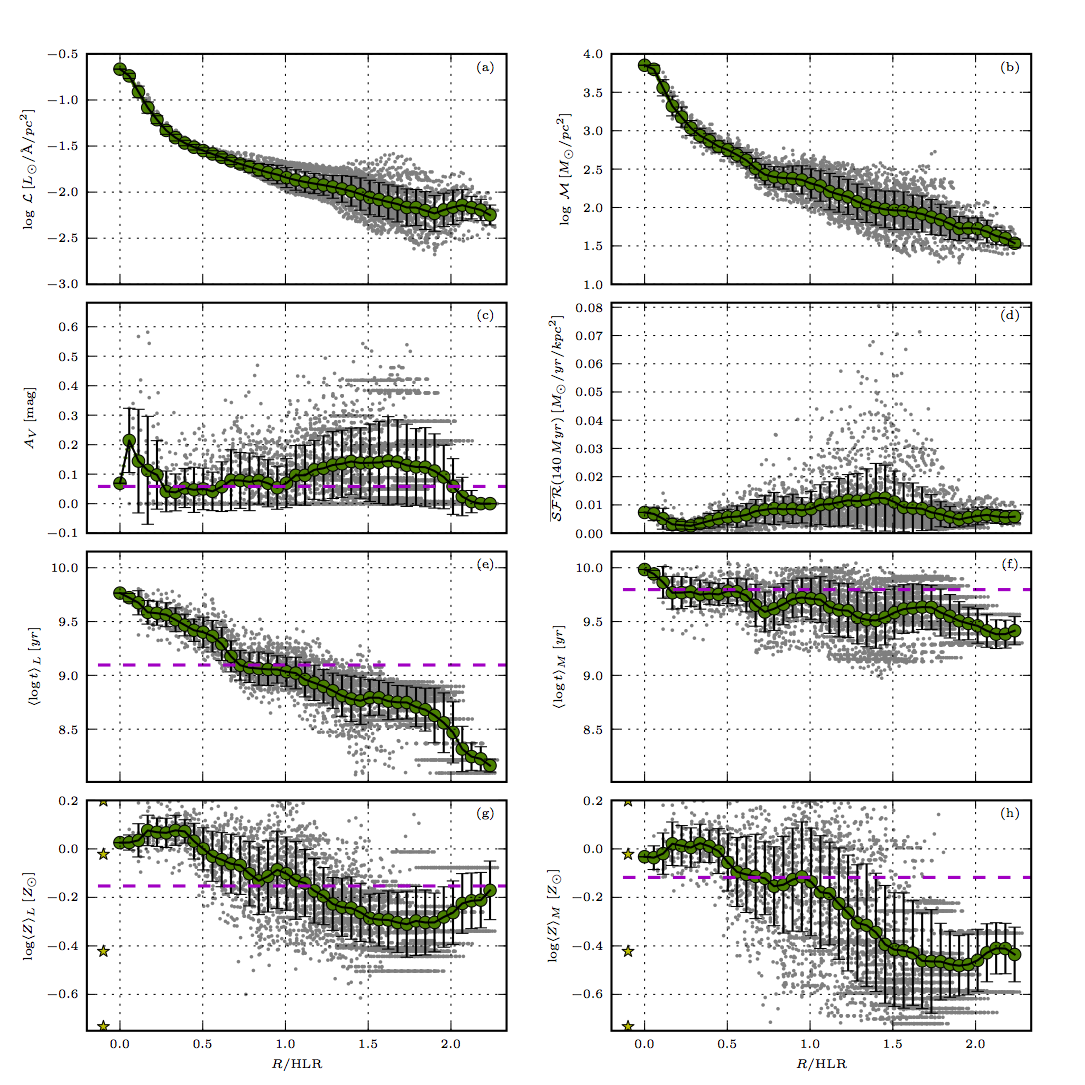}

\caption{Radial profiles of several properties. Grey points correspond to values
in individual $xy$ spaxels.
The green circles and  solid lines represent the
average of the plotted quantity, while the error bars represent the dispersion
in the $R$-bin. The dashed magenta horizontal line (only in the plots not
involving surface densities) marks the value derived from the \starlight\
analysis of the integrated spectrum, i.e. collapsing the $xy$ dimensions of the
datacube. The stars in the bottom panels mark the 4 metallicities in the base
models. }

\label{fig:radialProfiles} 
\end{figure*}

The information contained in 2D maps like the ones shown above can sometimes be
hard to absorb. Azimuthal averaging is a useful way to compress and help
digesting 2D data. \pycasso\ provides both circular and elliptical $xy$-to-$R$
conversion tensors. The examples below are based on a circular mapping.

For quantities like ${\cal M}$, one can think of two types of radial averaging.
The first is to add up all the stellar mass in $xy$ spaxels within a given
$R$-bin and then divide by the bin area (counting only non-empty spaxels). The
second is to average the surface density values for each spaxel directly.  The
same applies to, for instance, $\langle \log t \rangle_L$ (eq.\
\ref{eq:at_flux}): One can either add up the value of product $L_{txy} \times
\log t$ for all spaxels at a given $R$ and then divide by the corresponding
$\sum L_{xy}$, or else simply average the $\langle \log t \rangle_{L}$ values for each spaxel in
the $R$-bin. The first method, which we may call {\em area averaging},
effectively collapses the galaxy to 1D, but this kind of averaging cannot be
applied to quantities which do not involve light or mass, like $A_V$ and
$\sigma_\star$. For the sake of uniformity,  in this subsection we  chose the
second type of radial averaging (henceforth  {\em spaxel averaging}), noting
that the two methods almost always give nearly identical results.

Fig.\ \ref{fig:radialProfiles} shows several \pycasso\ products as a function of
$R$. The grey points represent values for individual spaxels. The solid line and green circles show
the mean value of the plotted quantity among the spaxels in the same $R$ bin, and the error bars map the corresponding dispersion. (As discussed in Paper II, actual error bars in radial profiles are much smaller due to the large number statistics.) Several of the features discussed  while describing the 2D images are clearly seen in these plots. The $\langle \log t \rangle$ and $\langle Z \rangle$ gradients, in particular, are cleanly depicted in these plots. 
As pointed out in Section \ref{sec:mean_ts_and_Zs}, means ages and metallicities increase $\sim$ simultaneously towards the nucleus, while $A_V$ does not change much, indicating that the classical degeneracies in stellar population studies are not playing a strong role in shaping the results, with the possible exception of the central regions. 
As in Fig.\ \ref{fig:kinematics}, horizontal stripes of gray points in panels c, e, f, g and h of Fig.\ \ref{fig:radialProfiles} correspond to spaxels in the same Voronoi zones. The stripes disappear in panels a, b and d because these extensive quantities were "dezonified" (cf.\ eq.\ \ref{eq:Mcor_xy}).

\subsection{1D temporal maps: evolutionary curves}

\label{sec:timeProfiles}

\begin{figure*}
\includegraphics[width=1.0\textwidth]{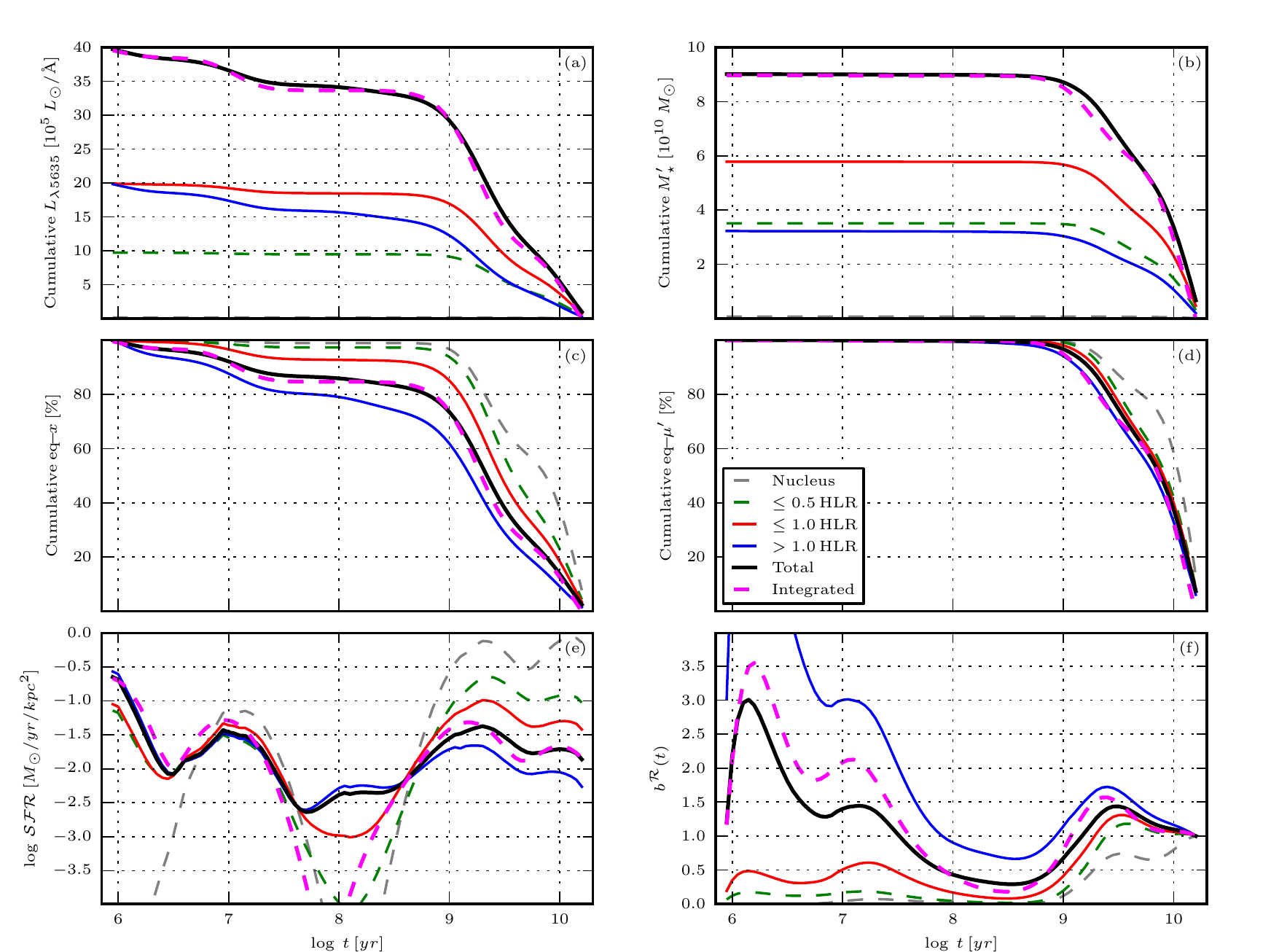}

\caption{Time evolution of light and mass representations of the SFH, as derived from
the \starlight\ fits. The discrete $x_{tZxy}$ and $\mu^\prime_{tZxy}$
"population vectors" involved in these curves were smoothed in $\log
t$, marginalized in $Z$ and integrated over the spatial regions indicated.
Dotted grey lines indicate the nucleus (central pixel). Green, red, and blue
lines correspond to $R \le 0.5$ HLR, $R \le 1$ HLR and $R> 1$ HLR regions,
respectively. The black line represents the whole galaxy evolution as
reconstructed from the sum of its parts, while (as in Fig.\
\ref{fig:radialProfiles}) the magenta dashed line is used to represent the
results obtained for the spatially integrated spectrum. }

\label{fig:timeProfiles} 
\end{figure*}

The main power of fossil record methods is to open the time domain, allowing
evolutionary studies. The caveat is that it does so with the typical logarithmic
resolution inherent to stellar evolution, but several studies 
 show that much can be learned about galaxy
evolution despite this fundamental constraint. Clearly, though, our 39 ages base
is highly overdimensioned, so some sort of SFH compression is necessary. As
discussed by Asari et al.\ (2007), age resolutions between 0.5 and 1 dex are
reasonable. We thus smooth the light and mass population vectors  with a
gaussian of FWHM $= 0.5$ dex in $\log t$.

Fig.\ \ref{fig:timeProfiles} plots a number of our synthesis products against
age. To keep an at least partial representation of the spatial information, we
plot evolutionary curves derived for different radial regions: The nucleus,
defined as the central pixel (plotted in grey dashed lines), $R \le 0.5$ (dashed green), $ \le 1$ (solid red), and $ > 1$ HLR (solid blue).

The top panels show the growth of luminosity (Fig.\ \ref{fig:timeProfiles}a) and
mass (b).\footnote{For clarity, the cumulative mass is computed with the total
mass which has participated in star formation, instead of the one corrected for
returned mass. In other words, we cumulate $M^\prime$ instead of $M$.} Among other things,
this plot illustrates the difference between light and mass, in the sense that
regions inside and outside $R = 1$ HLR do not have the same mass, despite having
(by definition) the same light. While HLR $= 4.8$ kpc, the radius containing
half of the mass is 3.4 kpc.  This is a direct
consequence of the stellar population gradients in this galaxy.
Gonz\'alez Delgado et al.\ (in prep) investigates the relation between light and mass based radii for different types of galaxies.


It is also seen that both light and mass grow at {\em different speeds for
different regions}. This is better appreciated in panels c and d, were
each growth curve is plotted on a 0 to 1 scale, with 1 representing the present
values (which tantamounts  to cumulating the equivalent $x_t$ and $\mu^\prime_t$
vectors for each region). The nucleus reached 80\% of its mass at $t = 8.5$ Gyr,
while the $R> 1$ HLR region did so later, at $t = 1.9$ Gyr. As shown by P\' erez et al.\ (2013), this inside-out ordering of the
mass assembly history applies to essentially all massive galaxies.

Fig.\ \ref{fig:timeProfiles}e  shows ${\cal SFR}(t)$. The SFR per unit area
decreases from the nucleus outwards through most of the time, but the trend is
reversed in the last $\sim 500$ Myr.  Notice that, because of the
smoothing, ${\cal SFR}(t)$ is now a continuous function of time, so that this
panel represent the {\em instantaneous} SFR, as opposed to the mass-over-time
definition in eq.\ \ref{eq:MoT}. Fig.\ \ref{fig:timeProfiles}f shows Scalos's $b$ parameter
for each region, which does use the running mean $\overline{\cal SFR}(t)$ of
eq.\ \ref{eq:MoT}. This plot helps interpreting Fig.\ \ref{fig:bScalo}, which
opens up the radial regions into full 2D maps, but compress the time axis by
stipulating fixed $t_{SF}$ time scales.

\subsection{The whole versus the sum of the parts}

\label{sec:TheWhole_x_SumOfTheParts}

One of the applications of CALIFA is to assess the effects of the lack of spatially resolved information on the derivation of physical properties and SFHs from integrated-spectra surveys. The horizontal dashed magenta lines in Fig.\ \ref{fig:radialProfiles} illustrate this point. They represent the results obtained from the analysis of the spectrum of  datacube {\em as a whole}, ie., adding upp all spaxels to  emulate an integrated spectrum. Qualitatively, one expects that this should produce properties typical of the galaxy zones as a whole. This expectation is borne out in Fig.\ \ref{fig:radialProfiles}, where one sees that the extinction, mean ages and metallicities marked by the dashed magenta line do represent an overall average of the spatially resolved values. In this particular example, they all match quite well the values at 1 HLR. The stellar masses obtained from the whole and the sum of the parts are also in excellent agreement, differening by just 1\%. Similarly, galaxy-wide luminosity and mass weighted mean ages and metallicities computed in these two ways agree to within 0.05 dex.


But can one derive the SFH of a galaxy out of an integrated spectrum? Fig.\
\ref{fig:timeProfiles} says that, at least in CALIFA 277, the answer is yes. The
dashed magenta and the black solid lines are very similar in all panels of this
figure. As before, the former represents the results obtained from the analysis
of the spatially compressed datacube, while the black curves
are computed adding the \starlight\ results for each spaxel. We are thus
comparing the whole against the sum of the parts, and they match.

At first sight, this may seem a trivial result, but it is not. Formally, one
only expects this to happen if $A_V$ is the same at all positions, a condition
which is approximately meet in CALIFA 277. In objects where dust has strong
spatial variations, the spectral fitting of the global spectrum with a single
$A_V$ will inevitably operate compensations, like increasing the age to
compensate for an understimated $A_V$ and vice-versa. Also, the combined effects
of population gradients and kinematics (e.g., disc population contributing more to
the wings of absorption lines than the ones from the bulge) are hard to predict.
Since we are presenting results for a single example galaxy, it remains to be
seen how general this ``coincidence'' is.

\subsection{Space $\times$ time diagrams: SFHs in 2D}

\begin{figure*}
\includegraphics[width=1.0\textwidth]{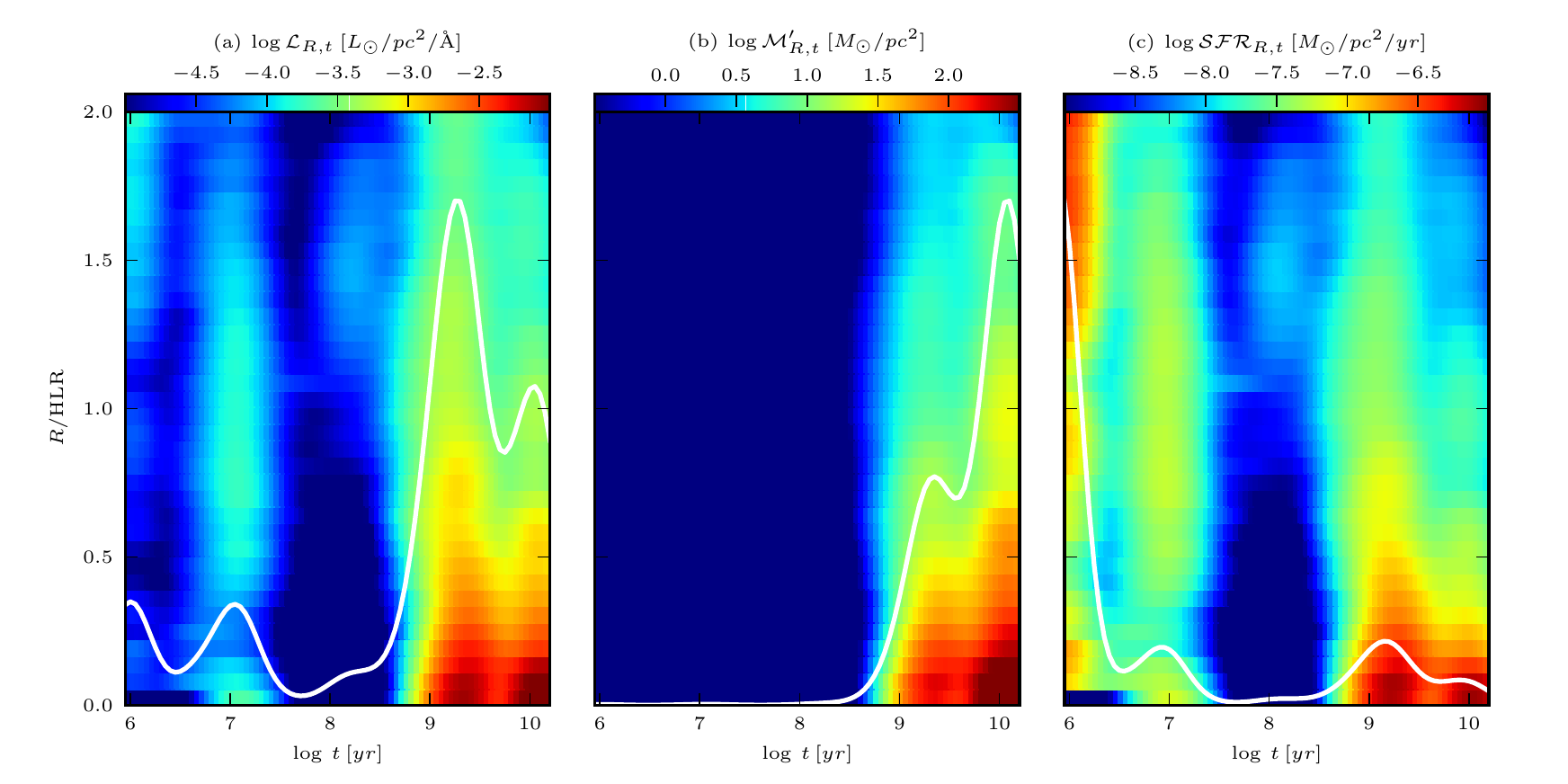}

\caption{$R$-$t$ diagrams showing the radially averaged distribution of light,
mass and SFR as a function distance from the nucleus and age. {\em Left:}
Luminosity at $\lambda = 5635$ \AA\ per unit area. $({\cal L}_{R,t})$. {\em
Middle:} Stellar mass formed per unit area $({\cal M}^\prime_{R,t})$. {\em
Right:} Time dependent SFR per unit area $({\cal SFR})$. The solid (white) lines
represent the sum over all spaxels for a given age.}
\label{fig:Rxt} 
\end{figure*}

As is evident at this point of the paper, one of the main challenges involved in the
analysis of the multi dimensional data built from the combination of the spatial
dimensions with the $t$ and $Z$ dimensions opened by population synthesis is how to
visualize the results. All examples shown so far project (or average over) two or more of these axes.

Fig.\ \ref{fig:Rxt} shows an attempt to visualize galaxy evolution as a function
of both space and time. The trick is to compress $xy$ into $R$, and collapse the
$Z$ axis, producing a radially averaged SFH$(R,t)$ map. The left panel shows the
luminosity density ${\cal L}$ at each radial position $R$ and for each age $t$.
As for the other panels in this figure, the original ${\cal L}_{tZyx}$ array
from which this map was derived was smoothed in $\log t$, marginalized over $Z$
and collapsed into one spatial dimension. Unlike in Fig.\
\ref{fig:radialProfiles}, we now use the area averaging method
discussed in Section \ref{sec:radialProfiles}, but the results do not depend
strongly on this choice. 

The ${\cal L}_{R,t}$ diagram shows that the light from old stars is concentrated
in the bulge, while the youngest stars shine $\sim$ equally at all $R$. Also, $t
\ga 1$ Gyr stars are more smoothly distributed in ages in the disc than in the
bulge, where they are concentrated in two populations, one at $\sim 2.5$ Gyr and
the other older than 10 Gyr. Fig.\ \ref{fig:Rxt}b shows the
formed stellar mass surface density as a function of $R$ and $t$. As usual, due
to the highly non-linear relation between light and mass in stars, the young
populations which show up so well in light  practically disappear when seen in
mass. Young stars reappear in the right panel, which shows the evolution in
time and space of the SFR surface density. As in Fig.\ \ref{fig:timeProfiles}g,
these are obtained from the instantaneous SFR at each spaxel, computed as in
Asari et al.\ (2007):

\begin{equation}
{\rm SFR}(t) = 
\frac{M^\prime_t}{\Delta t} =
\frac{\log e}{\Delta \log t}
\frac{M^\prime_t}{t}
\end{equation}

\noindent The age sampling is logarithmic ($\Delta \log t = $ constant), so
that the SFR is proportional to the stellar mass formed at $t$ divided by $t$.
The ${\cal SFR}_{R,t}$ and ${\cal M^\prime}_{R,t}$ diagrams therefore carry the same information; it is the $t^{-1}$ factor which makes them look so
different.

When interpreting these or any other plot involving age, one should always keep
in mind the logarithmic age resolution. For instance, taken at face value, the
spatially integrated SFR of CALIFA 277 in the last few Myr is almost equal to
its historical peak, $\sim 2.5$ Gyr ago (solid line in Fig.\ \ref{fig:Rxt}c).
However, these two peaks span {\em vastly different time intervals}, roughly by
a factor of Gyr/Myr = a thousand. The peak SFR around 2.5 Gyr ago, and indeed
even well before that (at $ > 10$ Gyr, when most of the mass was formed), were
surely  higher than the one we are now seeing at a few Myr, but cannot be
resolved in time. No fossil record method will ever be able to distinguish
bursts much shorter than their current age. These limitations are well known in
the field, but it is fit to recall them to avoid missinterpretation of the
results. 

Notwithstanding such age resolution issues, Fig.\ \ref{fig:Rxt} represents a
novel way of looking at galaxies both in time and space, and it shows clear
evidence in favor of an inside out growth scenario, in which the outer regions
assemble their mass at a slower pace and a over more extended period. This behavior is not unique to this galaxy. P\' erez et al (2012) find just the same in a study of the first 105  galaxies observed by CALIFA.

\subsection{Space vs.\ time "snapshots"}

\begin{figure}
\includegraphics[width=0.5\textwidth]{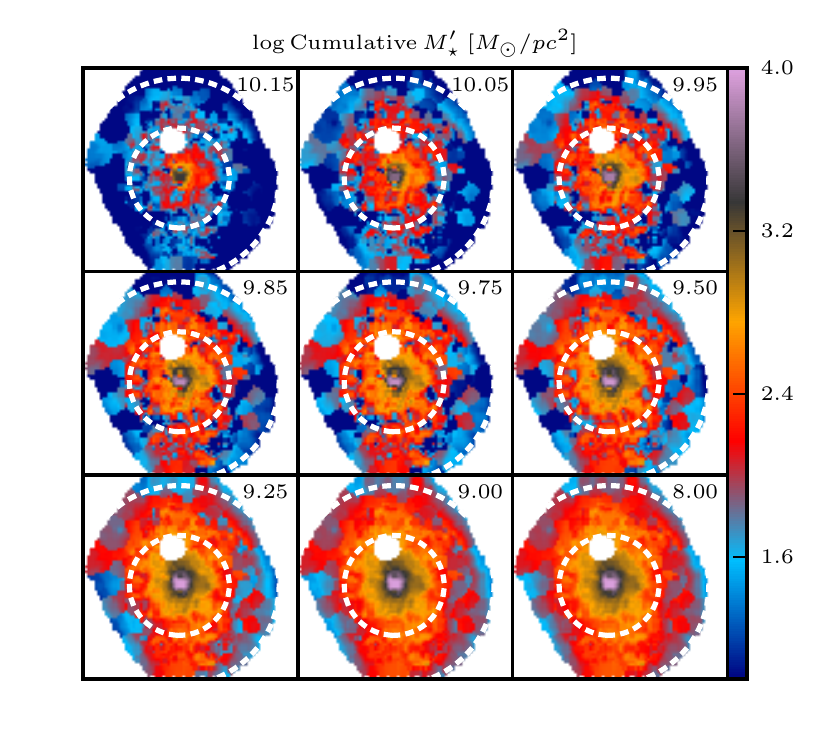}
\caption{"Pseudo snapshots" of the cumulative stellar mass surface density of CALIFA 277 at different ages: From $\log t$ [yr] $= 10.15$ to 8 (as labeled in each panel).}
\label{fig:mass_xy_x_t} 
\end{figure}

There is so much one can plot in a single image. To go beyond the $R$-$t$
diagrams of Fig.\ \ref{fig:Rxt}  one needs more dimensions than a sheet of paper
can accommodate. An alternative is shown in Fig.\ \ref{fig:mass_xy_x_t}, which
shows $t$-slices through the cumulative ${\cal M}^\prime_{xyt}$ cube. Unlike the
previous diagrams, which compressed the SFH in one way or another, these plots
reveal the full 2D nature of the mass assembly history of CALIFA 277.

Age  sequences like those in Fig.\ \ref{fig:mass_xy_x_t}  can be constructed for
several, but not all properties inferred though \starlight\ or any other fossil
method. One can, for instance, replace mass, by intrinsic luminosity, SFR or
stellar metallicity, in cumulative or differential form, but it is
obviously impossible to reconstruct maps of $A_V$, $\sigma_\star$, $v_\star$ as
a function of $t$. It is also worth pointing out that the panels in Fig.\
\ref{fig:mass_xy_x_t}  are {\em not} truly snapshots of a movie, ie., they are
not pictures of the galaxy as it appeared at different look-back times. Instead,
these are maps of where stars of a given age $t$ are located nowadays.

Simulators should take notice of these simple facts. IFS data plus fossil methods provide a rich, yet inevitably limited form of time-travel. Illustrative and beautiful as they are, movies of stars and gas particles moving through time and space will never be directly compared to anything observational. In other words, simulations should be convolved through this ``reality filter''. The observationally relevant predictions are the distribution of stellar ages and metallicities as a function of $xy$.

\subsection{Emission line maps}

\label{sec:EmLines}

\begin{figure}
\includegraphics[width=0.5\textwidth]{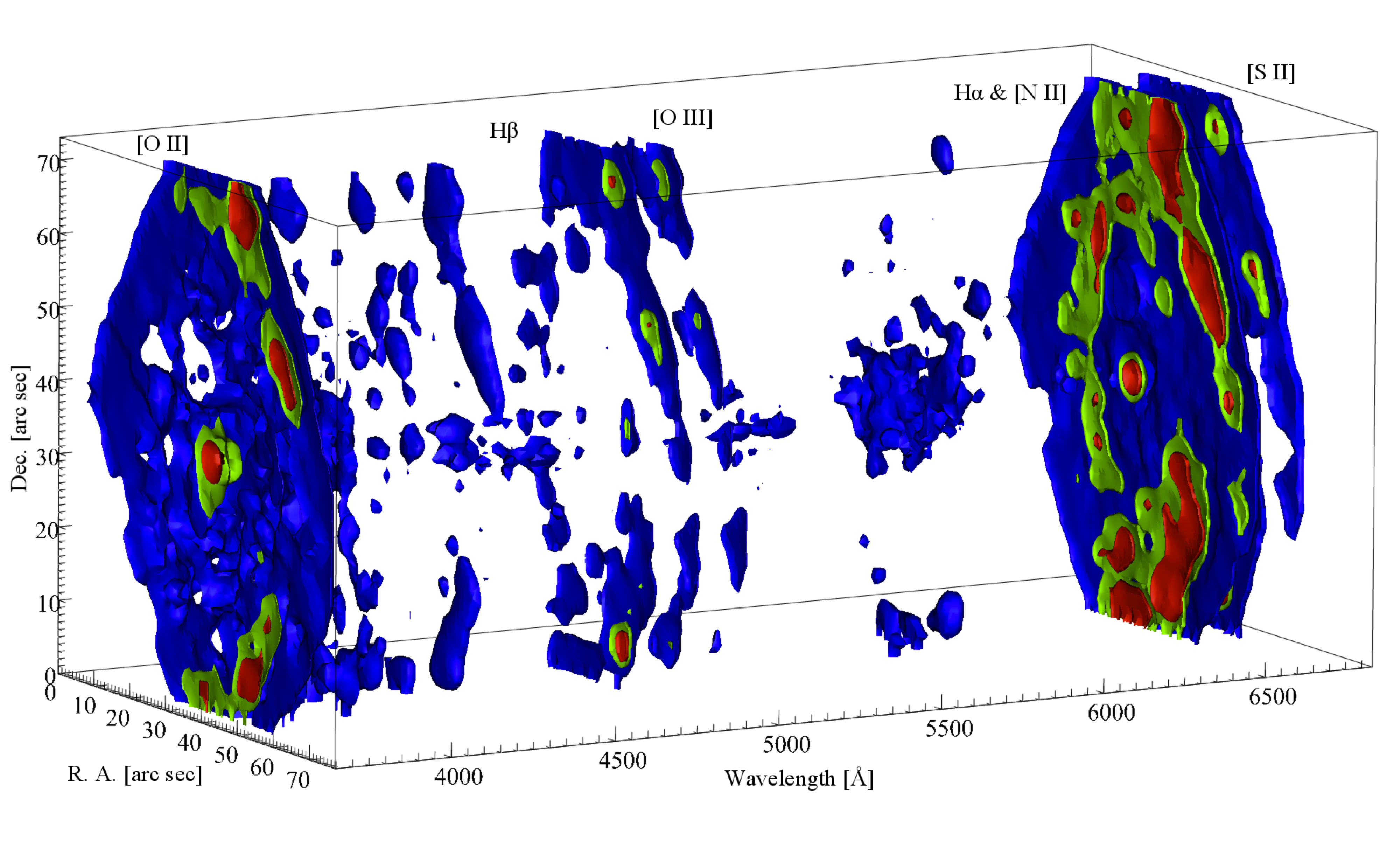} 

\caption{Residual spectral cube in a 3D rendering. }

\label{fig:EmLines} 
\end{figure}

Independent of the stellar population applications described throughout this
paper, the \starlight\ fits provide a residual spectrum $R_\lambda = O_\lambda - M_\lambda$ which is,
inasmuch as possible, free of stellar light. This is 
of great aid in
the measurement of emission lines, and hence in the derivation of a series of
diagnostics of the warm ISM properties, such as its kinematics, nebular
extinction and chemical abundance. The complementarity of nebular and stellar
analysis has been amply explored in studies of integrated spectra (eg, Cid
Fernandes et al.\ 2007; Asari et al.\ 2007; P\'erez-Montero et al.\ 2010; Garc\'{\i}a-Benito
\& P\'erez-Montero 2012), and the addition of two spatial dimensions to this
analysis holds great promise. Kehrig et al.\ (2012), for instance, combine \starlight\ analysis and
nebular emission to study the impact of post-AGB populations on the ionization
of the ISM in early type galaxies.

Fig.\ \ref{fig:EmLines} shows a  $R_{\lambda,xy}$ residuals cube, with some of
the main emission lines marked. The plot shows that emission lines are brighter towards the outside, in
particular over the spiral arms, as could be guessed from the distribution of
actively star-forming regions traced by our SFH analysis (e.g., Fig.\
\ref{fig:bScalo}). Despite its active nucleus, the
nuclear line fluxes are not particularly bright, implying a low luminosity AGN.

\section{Summary}

\label{sec:Conclusions}

This paper described a complete "analysis pipeline" which processes reduced IFS data through the \starlight\ spectral synthesis code. Data for the spiral galaxy NGC 2916 (CALIFA 277) were used to illustrate the journey from $(x,y,\lambda)$ datacubes to multi-dimensional maps of the synthesis products, focusing on the technical and methodological aspects. The three main steps in the analysis can be summarized as follows:

\begin{enumerate}

\item Pre-processing: This step comprises (a) construction of a spatial mask;
(b) adjustments in the flag and error spectra; (c) rest-framing and resampling;
(d) spatial binning. The last three stages were coded in a generic python
package ({\sc qbick}) which can run in a fully automated fashion. Spatial binning was
accomplished with an Voronoi tesselation scheme, tuned to produce zones with $S/N \ge 20$
spectra. In practice, because of the good quality of the data, no binning was
necessary within $R \la 1$ HLR. The issue of correlated errors in spatial
binning was discussed and an empirical recipe to correct for it was presented.
The output of these pre-processing steps are datacubes with fluxes, errors and
flags ready for a detailed $\lambda$-by-$\lambda$ spectral analysis, as well as
spatial masks, a zone map and other ancillary products.

\item \starlight\ fits and organization of the synthesis results: Spectral fits
for all spatial zones were performed using a base of SSP spectra from a
combination of MILES and Granada models, spanning ages from 1 Myr to 14 Gyr and
4 metallicities between 0.2 and 1.5 solar. The results were packed in a coherent
multi-layered FITS (or HDF5) file with the Python Califa Starlight Synthesis Organizer,
\pycasso.

\item Post-processing tools: \pycasso\ includes a reader module and a series of
analysis tools to perform operations like zone-to-pixel image reconstruction,
mapping  the spectral and stellar population properties derived by \starlight\
into multi-dimensional time, metallicity, and spatial coordinates, averaging in
spatial and temporal dimensions, manipulation of the stellar population arrays,
etc.

\end{enumerate}

Some of the products of this whole analysis are standard in IFS work, like the
kinematical field and emission line maps. The real novelty, of
course, resides in the spatially resolved stellar population products recovered
from the fossil record of galaxy evolution encoded in the datacubes. 
The 2D products range from maps of the stellar extinction and surface densities
of stellar mass to more elaborate products like mean ages and metallicities,
time averaged star formation rates and $b$-parameters. These quantities are all
normally used in the analysis of integrated spectra, but their application to
IFS data raises some conceptual issues, related to the unavoidable fact that not
all stars in a given spaxel were born there. On the other hand, one can use
spatially resolved data to construct diagnostics unaplicable to integrated data,
such as indices comparing the present and past SFR in different regions, or
comparing the ``present here'' with the ``present elsewhere''. These IFS-based
variations of the traditional $b$ parameter were shown to be useful to highlight
different aspects of the spatially resolved SFH.

1D profiles in temporal or radial coordinates were used as a means to help
interpreting the results. In our example galaxy, these compressed views of the
$(t,Z,x,y)$ manifold reveal clear age and metallicity negative gradients, as
well as spatially dependent mass growth speeds compatible with an inside-out
formation scenario. Finally, radius-age diagrams and ``snapshot'' sequences
were  introduced as further means of visualization.

Clearly, the application of spectral synthesis methods to IFS datacubes offers
new ways of studying galaxies and their evolution. Future communications will
use the tools laid out in this paper to explore the astrophysical implications
of the results for galaxies in the CALIFA survey.

\begin{acknowledgements} 
CALIFA is the first legacy survey being performed at Calar Alto. The CALIFA collaboration would like to thank the IAA-CSIC and MPIA-MPG as major partners of the observatory, and CAHA itself, for the unique access to telescope time and support in manpower and infrastructures.  We also thank the CAHA staff for the dedication to this project.

RCF thanks the hospitality of the IAA and the support of CAPES and CNPq. ALA acknowledges support from INCT-A, Brazil. BH gratefully acknowledges the support by the DFG via grant Wi 1369/29-1.  Support from the Spanish Ministerio de Economia y Competitividad, through projects AYA2010-15081 (PI RGD), AYA2010-22111-C03-03 and AYA2010-10904E (SFS), AYA2010-21322-C03-02 (PSB) and the Ram\'on y Cajal Program (SFS, PSB and JFB), is warmly acknowledged. We also thank the Viabilidad , Dise\~no , Acceso y Mejora funding program, ICTS-2009-10, for funding the data acquisition of this project.

\end{acknowledgements}

\end{document}